\documentclass[12pt]{article}

\usepackage{amssymb,amsfonts,amsthm}
\usepackage{amsmath}
\usepackage{bm}             
\usepackage[mathscr]{eucal}
\usepackage{cancel}
\usepackage{wasysym}

\def\be{\begin{equation}}
\def\ee{\end{equation}}
\def\bea{\begin{eqnarray}}
\def\eea{\end{eqnarray}}

\def\bdelta{\mbox{\boldmath $\delta$}}

\def\la{\langle}
\def\ra{\rangle}
\def\hsp5{\hspace{5mm}}

\theoremstyle{remark}

\newcommand{\cH}{\mathcal{H}}

\newcommand{\sfrac}[2]{{\textstyle{#1\over#2}}}

\title{\sc Second order density perturbations for dust cosmologies}

\begin{document}

\author{ \\
{\Large\sc Claes Uggla}\thanks{Electronic address:
{\tt claes.uggla@kau.se}} \\[1ex]
Department of Physics, \\
University of Karlstad, S-651 88 Karlstad, Sweden
\and \\
{\Large\sc John Wainwright}\thanks{Electronic address:
{\tt jwainwri@uwaterloo.ca}} \\[1ex]
Department of Applied Mathematics, \\
University of Waterloo,Waterloo, ON, N2L 3G1, Canada \\[2ex] }

\date{}
\maketitle

\begin{abstract}

We present simple expressions for the relativistic first and second order
fractional density perturbations for FL cosmologies with dust, in four
different gauges: the Poisson, uniform curvature, total matter and
synchronous-comoving gauges. We include a cosmological constant and arbitrary spatial
curvature in the background.  A distinctive feature of our approach is our
description of the spatial dependence of the perturbations using a canonical
set of quadratic differential expressions involving an arbitrary spatial
function that arises as a conserved quantity. This enables us to unify,
simplify and extend previous seemingly disparate results. We use the
primordial matter and metric perturbations that emerge at the end of the
inflationary epoch to determine the additional arbitrary spatial function
that arises when integrating the second order perturbation equations. This
introduces a non-Gaussianity parameter into the expressions for the second
order density perturbation. In the special case of zero spatial curvature we
show that the time evolution simplifies significantly, and requires the use
of only two non-elementary functions, the so-called growth supression factor
at the linear level, and one new function at the second order level. We
expect that the results will be useful in applications, for example, studying
the effects of primordial non-Gaussianity on the large scale structure of the
universe.

\end{abstract}

\centerline{\bigskip\noindent PACS numbers: 04.20.-q, 98.80.-k, 98.80.Bp,
98.80.Jk}

\section{Introduction}

The increasingly accurate observations of the cosmic microwave background
(CMB) and the large scale structure (LSS) of the universe that are becoming
available are stimulating theoretical developments in cosmology. The analysis
of these observations is based on cosmological perturbation theory, with the
bulk of the work to date using linear perturbations of Friedmann-Lemaitre
(FL) cosmologies.  However, the study of possible deviations from linearity,
for example, how primordial non-Gaussianity affects the anisotropy of the CMB
and the LSS, necessitates the use of nonlinear perturbations (see for
example~\cite{ baretal04b, baretal10b,pitetal10}).

Much of the theoretical work on nonlinear perturbations has dealt with flat
FL cosmologies with dust (a matter-dominated universe, see for
example~\cite{matetal98,baretal05}) and more recently also with a
cosmological constant (a $\Lambda$CDM universe, see for
example~\cite{tom05}--\cite{bruetal14}). One aspect of the work is to provide
an expression for the second order fractional density perturbation
${}^{(2)}\!{\bdelta}$  which is needed  when analyzing  observations of the
LSS, and this is the focus of the present paper.

The analysis of the LSS is based on galaxy redshift surveys, which record the
fluctuation of the number count of galaxies on the past light cone of the
observer as a function of redshift and angular direction. On the other hand
the fractional density perturbation, calculated using cosmological
perturbation theory, is based on constant time slices in the Universe,\emph{
i.e.} it depends on a choice of temporal gauge. Thus in relating observations
to theory one has to take into account both gauge effects and lightcone
effects. This has been done in detail in linear perturbation theory (see, for
example,~\cite{chalew11}-\cite{yooetal12}), and more recently in second order
theory~\cite{beretal14}. These papers derive a formula that relates the
observed fractional galaxy number overdensity as a function of redshift $z$
and direction ${\bf n}$ to the fractional density perturbation as a function
of spatial position and time in an appropriate gauge. In the present paper we
focus on one aspect of this whole process by deriving simple explicit
expressions for the relativistic first and second order fractional density
perturbations for FL cosmologies with dust, and investigating how the choice
of gauge affects the structure of the expressions.

In a recent paper~\cite{uggwai13}, here referred to as UW, we derived a
general expression for ${}^{(2)}\!{\bdelta}$ using the Poisson gauge, also
including the effects of spatial curvature, but subject to the restriction
that the perturbation at linear order is purely scalar.\footnote{For the
motivation for imposing this restriction, we refer to~\cite{pitetal10}, page
4.} This expression for ${}^{(2)}\!{\bdelta}$ contains the integral of a
complicated quadratic source term involving two arbitrary spatial functions,
which makes it difficult to obtain a clear physical understanding. If one
assumes that the decaying mode of the scalar perturbation at the linear order
is zero, however, then the scalar perturbation depends on only one such
function. In this case we showed that the temporal and spatial dependence of
${}^{(2)}\!{\bdelta}$ can be displayed explicitly, in a form that provides
direct physical insight.

In the present paper we investigate what effect the choice of gauge has on
the form of ${}^{(2)}\!{\bdelta}$. Specifically, we first derive a new
expression\footnote{This expression differs from the one in UW referred to
above in the way the time dependence is represented, which facilitates the
subsequent analysis.} for  ${}^{(2)}\!{\bdelta}$ in the Poisson gauge by
solving the second order perturbation equations as given in UW. We then use
the change of gauge formulas in appendix~\ref{sec:transf} to calculate
${}^{(2)}\!{\bdelta}$ in three other commonly used gauges: the uniform
curvature, the total matter\footnote{The total matter gauge is referred to by
Hwang and coworkers as the comoving gauge (see for example~\cite{hwanoh06}
and~\cite{hwaetal12}). We also used this terminology in an earlier
paper~\cite{uggwai12}. Here, however, we have chosen to follow the
conventions and nomenclature of Malik and Wands~\cite{malwan09}, see their
sections 7.4 and 7.5.} and the synchronous-comoving gauges.  Before
continuing, we digress briefly to motivate our choice of gauges.\footnote{We
give the definition of the gauges in appendices~\ref{sec:poisson}
and~\ref{sec:synch}.} The two most commonly used gauges in cosmological
perturbation theory are the synchronous-comoving gauge and the Poisson gauge.
For linear perturbations of dust cosmologies the total matter gauge is the
same as the synchronous-comoving gauge. For second order perturbations,
however, this is no longer the case. Hence we chose the total matter gauge as
one of the four gauges. Finally we decided to use the uniform curvature gauge
because we noticed that for this gauge the second order fractional density
perturbation has an interesting property, namely \emph{the superhorizon part
is a conserved quantity}.\footnote{This generalizes the known result for the
linear fractional density perturbation  in the uniform curvature gauge.} We
note that this result plays a key role in the process of simplifying the
expressions for the second order fractional density perturbation that we give
in this paper.

We will use the notation ${}^{(2)}\!{\bdelta}_{\bullet}$ for the fractional
density perturbation, where the bullet identifies the gauge. Our first main
result is that ${}^{(2)}\!{\bdelta}_{\bullet}$ has the following common
structure for the Poisson, the uniform curvature and the total matter gauges:
\be \begin{split} {}^{(2)}\!{\bdelta}_{\bullet} &= \quad \underbrace{A_{1,{\bullet}}\,\zeta^2 +
A_{2,{\bullet}}\,{\cal D}(\zeta)}_{\text{the super-horizon part}} \\
&\quad  + \underbrace{\sfrac23m^{-2}xg\left[A_{3,{\bullet}}\,({\bf D}\zeta)^2 + A_{4,{\bullet}}\,{\bf
D}^2{\cal D}(\zeta) + A_{5,{\bullet}}\,{\bf D}^2 \zeta^2 \right]}_{\text{the post-Newtonian part}}  \\
&\quad  + \underbrace{\sfrac49m^{-4}x^2g^2\left[{\mathcal B}_3{\bf D}^2({\bf D}\zeta)^2 +
{\mathcal B}_4{\bf D}^4{\cal D}(\zeta)  \right]}_{\text{the Newtonian part}}.
\end{split}  \label{delta_2}  \ee
Here $m$ is a constant and the coefficients $A_{i,{\bullet}}$, $i=1,\dots,5$,
${\mathcal B}_3$ and ${\mathcal B}_4$ are functions of the scale factor $x$,
while the spatial dependence is determined by the conserved quantity
$\zeta(x^i)$, which  appears in seven quadratic differential
expressions:
\be  \zeta^2, \quad {\cal D}(\zeta), \quad ({\bf D}\zeta)^2, \quad {\bf
D}^2{\cal D}(\zeta), \quad {\bf D}^2 \zeta^2, \quad {\bf D}^2({\bf
D}\zeta)^2, \quad {\bf D}^4{\cal D}(\zeta).  \label{quad_zeta}   \ee
The spatial differential operators in~\eqref{delta_2} and~\eqref{quad_zeta}
are defined in appendix~\ref{sec:opdef}. We will refer to the group of terms
in~\eqref{delta_2} whose differential expressions have zero weight in the
spatial differential operator ${\bf D}_i$ as \emph{the super-horizon part}.
The intermediate group of terms having weight two in ${\bf D}_i$ (coefficient
$m^{-2}$) is referred to as \emph{the post-Newtonian part}. Finally, we refer
to the group of terms in~\eqref{delta_2} having weight four in ${\bf D}_i$
(coefficient $m^{-4}$) as \emph{the Newtonian part}.

The common structure for ${}^{(2)}\!{\bdelta}_{\bullet}$ in these three
gauges is due to the fact that they all use the same spatial gauge. The
differences thus depend on different temporal gauges, which affect the
coefficients $A_{i,{\bullet}}$, $i=1,\dots,5$, but not ${\mathcal B}_3$ and
${\mathcal B}_4$. On the other hand, the synchronous-comoving gauge uses a different
spatial gauge, which has the effect of adding a term to the Newtonian part
with spatial dependence given by $({\bf D}^2\zeta)^2$, thereby adding a new
quadratic differential expression to the set~\eqref{quad_zeta}. For this
reason we will treat this case separately in the paper.

The evolution of ${}^{(2)}\!{\bdelta}$ in the Poisson gauge is determined by
eight functions of time that are written as integrals, and these are the main
source of the complexity of the expression. Our second main result is to show
that if the background spatial curvature is zero, then seven of these
integral functions can be written in an explicit form. This fact  enables us
to give simple expressions for ${}^{(2)}\!{\bdelta}$ in all four gauges under
consideration when the spatial geometry is flat.

The outline of the paper is as follows. In the next section we give a unified
expression for the  first order fractional density perturbation in the four
gauges. In section~\ref{Sec:secgen} we derive the corresponding second order
results, and address the issue of initial conditions. Section~\ref{sec:K0}
deals exclusively with the spatially flat case $K=0$. We show that the time
dependence simplifies significantly, and then give a detailed comparison with
previous work dealing with this case. Section~\ref{Sec:concl} contains the
concluding remarks. Finally, in appendix~\ref{sec:opdef} we define the
various spatial differential operators and in appendix~\ref{sec:transf} we
derive the transformation laws that relate the density perturbations for the
four gauges under consideration

\section{The density perturbation at first order}

The background cosmology is an  FL universe with scale factor $a$, Hubble
scalar $H$ and curvature parameter $K$, containing dust with background
matter density ${}^{(0)}\!\rho_m$ and a cosmological constant $\Lambda$. We
introduce the usual density parameters:\footnote{We use units such that $8\pi
G = 1 = c$.}
\be \Omega_m:=\frac{^{(0)}\!\rho_m}{3H^2}, \qquad \Omega_k:=- \frac{K}{{\cal
H}^2}, \qquad \Omega_{\Lambda}:=\frac{\Lambda}{3H^2},  \label{omega_def}  \ee
which satisfy
\be \Omega_m + \Omega_k + \Omega_{\Lambda}=1. \label{omega_sum} \ee
We use the dimensionless scale factor $x:=a/a_0$, normalized at some
reference epoch $t_0$, as time variable.\footnote {When $t_0$ is the present
time, $x = (1+z)^{-1}$, where $z$ is the redshift. We note that $x$ is
related to the conformal time $\eta$, which we shall sometimes use, according
to $\partial_{\eta}={\cal H}x\partial_x.$} The conservation law shows that
$a^3\,{}^{(0)}\!\rho_m$ is constant, which we write in terms of $\Omega_m$
and the dimensionless Hubble scalar ${\cal H}:=aH$ as
\be {\cal H}^2 x\,\Omega_m=m^2.  \label{omega_m} \ee
Setting $x=1$ shows that the constant $m$ is given by $m^2={\cal H}_0^2
\,\Omega_{m,0}$. It follows that
\be {\cal H}^2={\cal H}^2_0\,( \Omega_{\Lambda,0}\,x^2 + \Omega_{k,0} + \Omega_{m,0}\, x^{-1}). \label{calH} \ee
Equations~\eqref{omega_def}-\eqref{calH} determine ${\cal H}$, $\Omega_m$,
$\Omega_k$ and $\Omega_{\Lambda}$ explicitly in terms of $x$.

The gauge invariants\footnote{These gauge invariants are introduced in
appendix~\ref{sec:transf}. Our strategy for working with gauge invariants is
described in the first paragraph of that appendix.} that describe the  scalar
linear perturbations of the metric and matter in the Poisson gauge are the
Bardeen potentials ${}^{(1)}\!\Psi_{\mathrm p}$ and ${}^{(1)}\!\Phi_{\mathrm
p}$, the velocity potential ${}^{(1)}\!{\bf v}_{\mathrm p}$  and the
fractional density perturbation ${}^{(1)}\!{\bdelta}_{\mathrm p}$. In the
special case when the decaying mode of the scalar perturbation is set to zero
we have the following expressions:\footnote{See equations (19b) and (30)-(32)
in UW.}
\begin{subequations} \label{p1}
\begin{align} {}^{(1)}\!{\Psi}_{\mathrm p} &={}^{(1)}\!{\Phi}_{\mathrm p}= g(x)\zeta(x^i), \qquad
 {\cal H}{}^{(1)}\!{\bf v}_{\mathrm p}=
-\sfrac23 \Omega_m^{-1} fg\,\zeta,  \label{psi_p} \\
{{}^{(1)}\!\bdelta}_{\mathrm p} &= -2\Omega_m^{-1}(f+\Omega_k) g\zeta
+\sfrac23  m^{-2}xg\,{\bf D}^2 \zeta, \label{del_p1}
\end{align}
\end{subequations}
where\footnote{We note in passing that the function $g(x)$, defined up to a
constant factor, is sometimes referred to as the growth suppression factor.
See for example~\cite{baretal06}, in the text following equation (2.3).  Our function
$f(x)$ equals their function $f(\eta)$ in equation (2.9), on noting that
$g'(\eta)={\cal H}x \partial_x g(x)$.}
\be  \label{fg}  g(x):= \sfrac32 m^2\frac{{\mathcal H}}{x^2}\int_0^x \frac{d{\bar x}}{{\mathcal H}({\bar x})^3},
\qquad  f(x):=1 + g^{-1}{x\partial_x g}. \ee
Here the arbitrary spatial function $\zeta(x^i)$ equals the
conserved quantity that we introduced in~\cite{uggwai12},
denoted by $\zeta_{\mathrm v}$, which can be written in the
form\footnote{Referring to UW, use equations (11) and (12) to rewrite equation
(21).}
\be \label{zeta}  \zeta_{\mathrm v} = ( 1 + \sfrac23\, \Omega_k
\Omega_m^{-1}){}^{(1)}\!\Psi_{\mathrm p} - {\cH}{}^{(1)}\!{\bf v}_{\mathrm
p}. \ee
We emphasize that $\zeta_{\mathrm v}$ \emph{is exactly conserved for dust}
(see~\cite{hwanoh99}, equation (B2), and~\cite{uggwai12} equations
(69)-(71)).  The fact that $\zeta=\zeta_{\mathrm v}$ was established
in UW (see equations (21)-(24)). We note in passing that if the background
spatial curvature is zero ($\Omega_k=0$) then $\zeta_{\mathrm v}$ is the
comoving curvature perturbation, often denoted by ${\mathcal R}$ (see~\cite{malwan09},
equation (7.46) in conjunction with (7.6) and (7.8)).

We can use~\eqref{zeta} to simplify the expression for
${}^{(1)}\!{\bdelta}_{\mathrm p}$, as follows.  On substituting for
${}^{(1)}\!\Psi_{\mathrm p}$ and ${}^{(1)}\!{\bf v}_{\mathrm p}$
from~\eqref{psi_p} into~\eqref{zeta} the function $\zeta$ cancels as a common
factor and we obtain the following algebraic constraint relating $f$ to $g$:
\be   (3\Omega_m +2\Omega_k + 2f)g = 3\Omega_m.  \label{fg2}  \ee
Using this we can simplify the expression~\eqref{del_p1} to obtain:
\be {}^{(1)}\!{\bdelta}_{\mathrm p} =  -3(1-g) \zeta +\sfrac23 m^{-2}xg\,{\bf
D}^2\zeta. \label{bdelta_p1} \ee
By combining this equation with the transformation
formulas~\eqref{ptoc1},~\eqref{ptov1} and~\eqref{bdelta_2s} in
appendix~\ref{sec:transf}, we can give a unified expression for
${}^{(1)}\!{\bdelta}_{\bullet}$, the gauge invariant associated with the
first order density perturbation in the four gauges that we are considering
in this paper:
\begin{subequations}\label{bdelta_1}
\be {}^{(1)}\!{\bdelta}_{\bullet} = A_{\bullet}\zeta + \sfrac23 m^{-2}xg\,{\bf D}^2 \zeta,\ee
where
\be A_{\mathrm p} =-3(1-g), \qquad A_{\mathrm c} =-3,   \qquad A_{\mathrm v}
= A_{\mathrm s}= -2\Omega_m^{-1}\,\Omega_k\,g. \ee
\end{subequations}
Here the subscripts $_{\mathrm p}$, $_{\mathrm c}$, $_{\mathrm v}$ and $_{\mathrm s}$
identify the Poisson, uniform curvature, total matter and synchronous-comoving gauges, respectively.

The first term of ${}^{(1)}\!{\bdelta}_{\bullet}$ in~\eqref{bdelta_1}, which
has zero weight in the spatial differential operator ${\bf D}_i$ is referred
to as \emph{the super-horizon term}, while the second term, which has weight
two is referred to as \emph{the Newtonian term}. We note that the
super-horizon term is gauge-dependent, while the Newtonian part is
gauge-independent for these four gauges.\footnote{Clearly the Newtonian part
is not \emph{universally} gauge-independent since for the uniform density
gauge it is zero.}

Finally the form of the super-horizon term in ${}^{(1)}\!{\bdelta}_{\mathrm
c}$, as given by~\eqref{bdelta_1}, deserves comment: it is independent of
time. This is to be expected since the quantity
$\zeta_{\rho}:=-\sfrac13{}^{(1)}\!{\bdelta}_{\mathrm c}$ satisfies the
"conservation law" $\partial_{\eta} \zeta_{\rho}=\sfrac12 {\bf D}^2 {\bf
v}_{\mathrm p}$, which suggests that ${}^{(1)}\!{\bdelta}_{\mathrm c}$ will
be constant in a regime in which spatial derivatives can be
neglected.\footnote{See~\cite{uggwai12}, equations (65) and (66) specialized
to a barotropic perfect fluid (${\bar {\Gamma}}=0$). See
also~\cite{wandsetal00}, equation (18) in conjunction with (7)-(9).}

The expressions for the gauge invariants at first order given by
equations~\eqref{p1},~\eqref{fg} and~\eqref{zeta} provide the foundation for
generalizing to second order perturbations. In particular the
constraint~\eqref{fg2} will be used frequently in simplifying the expressions
for the second order fractional density perturbation.

\section{The density perturbation at second order}\label{Sec:secgen}

In this section we derive explicit expressions for the second order
fractional density perturbation ${}^{(2)}\!{\bdelta}_{\bullet}$ in the Poisson,
uniform curvature, total matter and synchronous-comoving gauges,
subject to the following restrictions: i) the perturbations at linear order
are purely scalar, and ii) the decaying mode of the scalar perturbation is
zero.

\subsection{The Poisson gauge}

The gauge invariants\footnote{These gauge invariants are introduced in
appendix~\ref{sec:transf}.} that describe the  scalar second order
perturbations of the metric and matter in the Poisson gauge are the Bardeen
potentials ${}^{(2)}\!\Psi_{\mathrm p}$ and ${}^{(2)}\!\Phi_{\mathrm p}$, the
velocity potential ${}^{(2)}\!{\bf v}_{\mathrm p}$  and the fractional
density perturbation ${}^{(2)}\!{\bdelta}_{\mathrm p}$. The governing equations that
determine these gauge invariants in the case of dust cosmologies are given in
a concise form in equations (14)-(16) in UW. A particular solution for
${}^{(2)}\!\Psi_{\mathrm p}$ is given by equations (26) in UW:
\begin{subequations} \label{Psi2.bbS}
\be \label{Psi2_gen}  {}^{(2)}\!\Psi_{\mathrm p}(x,x^i)=\frac{{\mathcal H} }{x^2}
\int_{0}^x \frac{{\bar x}\,\Omega_m}{\cal H}\,{\mathbb S}({\bar x},x^i) d{\bar x}, \ee
where
\be \label{bbS_gen} {\mathbb S}(x,x^i):= m^{-2}\int_{0}^{ x} {\mathcal
S}({\tilde x},x^i) d{\tilde x}, \ee
\end{subequations}
and the source term ${\mathcal S}$ is given by equation (70a) in
UW.\footnote{We have modified equations (26) in UW by using the constraint
$x{\cal H}^2 \Omega_m=m^2$ to replace ${\cal H}^2$ in the first integral and
have rescaled ${\mathbb S}$ by a factor of $m^{-2}$. In addition we choose
$x_{initial}=0$, which is possible since we have set the decaying mode to
zero.}

We can  obtain a simple expression for the time
derivative of $ {}^{(2)}\!\Psi_{\mathrm p}$
by differentiating~\eqref{Psi2_gen} with respect to $x$
and using~\eqref{defq} and~\eqref{q_dust}:
\be \partial_x(x {}^{(2)}\!\Psi_{\mathrm p})=
-(\sfrac32\Omega_m+\Omega_k )  {}^{(2)}\!\Psi_{\mathrm p} +\Omega_m\,\mathbb S.
\label{partial_psi} \ee
On using this equation, the second order governing equations, given in UW as
equations (15) and (16),\footnote{The gauge invariants with symbols $\Psi,
{\bf v}$ and ${\bdelta}$ all refer to the Poisson gauge, which we indicate
here with a subscript $_{\mathrm p}$. Also note that $(\partial_{\eta} +{\cal
H}){}^{(2)}\!\Psi = {\cal H}x\partial_x(x{}^{(2)}\!\Psi)$ on changing time
derivative.} lead directly to the following expressions for
${}^{(2)}\!{\bdelta}_{\mathrm p}$ and ${}^{(2)}\!{\bf v}_{\mathrm p}$:
\begin{subequations} \label{delta2.v2}
\begin{align}
{}^{(2)}\!{\bdelta}_{\mathrm p} &=\sfrac23m^{-2}{\bf D}^2(x {}^{(2)}\!\Psi_{\mathrm p}) +
3{}^{(2)}\!\Psi_{\mathrm p} -2{\mathbb S} +{\bf S}_{\delta},  \label{delta2_p} \\
{\cal H}{}^{(2)}\!{\bf v}_{\mathrm p} &=(1+\sfrac23 \Omega_k\, \Omega_m^{-1}) {}^{(2)}\!\Psi_{\mathrm p} -
\sfrac23{\mathbb S} +{\bf S}_{v},   \label{v2_p}
\end{align}
\end{subequations}
where
\begin{subequations}\label{source_dv}
\begin{align}
{\bf S}_{\delta}&=\,{\mathcal S}_{\mathbb D} +3{\mathcal S}_{\mathbb V} - 2({\bf D}^{(1)}\!{\bf v}_{\mathrm p})^2, \\
{\bf S}_{v}&={\mathcal S}_{\mathbb V } -  2{\cal S}^i\left[({}^{(1)}\!{\bdelta}_{\mathrm p} -
{}^{(1)}\!\Psi_{\mathrm p}){\bf D}_i( {\cal H}{}^{(1)}\!{\bf v}_{\mathrm p})\right], \label{source_dv2}
\end{align}
\end{subequations}
and the source terms ${\mathcal S}_{\mathbb D}$ and ${\mathcal S}_{\mathbb
V}$ are given by equations (70b) and (70c) in UW. For the reader's
convenience we quote these expressions:\footnote{We have replaced the scalar
${\cal A}$ by ${\cal A}=3{\cal H}^2\Omega_m$, using equation (12) in UW.}
\begin{subequations}  \label{source}
\begin{align}
{\cal S}_{\mathbb D} &= \sfrac23({\cal H}{\Omega_m})^{-1}[4({\bf D}^2+3K)\Psi_{\mathrm p}^2 -5({\bf D}\Psi_{\mathrm p})^2 ] +
6 {\cal S}^i(\Psi_{\mathrm p} {\bf D}_i{{\cal H}{\bf v}_{\mathrm p}}) + \sfrac92\Omega_m ({\cal H}{\bf v}_{\mathrm p}) ^2, \\
{\cal S}_{\mathbb V} &= \sfrac23{\Omega_m}^{-1}\left[\Psi_{\mathrm p}^2 + 4{\cal D}(\Psi_{\mathrm p})\right] +
 2\left[{\cal S}^i({\cal H}{{\bf v}_{\mathrm p}}{\bf D}_i\Psi_{\mathrm p}) +2{\cal D}({\cal H}{{\bf v}_{\mathrm p}}) \right]. \label{S_V}
\end{align}
\end{subequations}
We note that the mode extraction operator ${\cal S}^i$, which is defined
by~\eqref{modeextractop}, satisfies ${\cal S}^i {\bf D}_i f=f$, for any
spatial function $f$, \emph{i.e.} ${\cal S}^i$ is the inverse operator of
${\bf D}_i$. For brevity we have omitted the superscript $^{(1)}$ on the
gauge invariants on the right side of equations~\eqref{source}.

The next step is to obtain explicit expressions for ${}^{(2)}\!\Psi_{\mathrm
p}$, ${}^{(2)}\!{\bdelta}_{\mathrm p}$ and ${}^{(2)}\!{\bf v}_{\mathrm p}$ by
substituting the first order solution~\eqref{p1} into
equations~\eqref{Psi2.bbS} and~\eqref{delta2.v2}. We begin by following UW in
obtaining an explicit expression for the source term ${\mathcal S}(x,x^i)$
in~\eqref{bbS_gen}. The result is\footnote{UW, equations (74)-(75). We have
rescaled the $T_A$ by multiplying by $m^{-2}$ and have used
equations~\eqref{omega_sum} and~\eqref{omega_m} in this paper.}
\be \label{S_spec}   {\mathcal S}(x,x^i)= m^2\left(T_1\zeta^2+
T_2{\cal D}(\zeta) + m^{-2}\left[T_3({\bf D}\zeta)^2 +
T_4{\bf D}^2 {\cal D}(\zeta)\right]\right),   \ee
where
\begin{subequations} \label{T_A}
\begin{align}
T_1(x) &:= (x\,\Omega_m)^{-1}g^2((f-1)^2- 4\Omega_k), \\
T_2(x) &:=  - 8(x\,\Omega_m)^{-1}g^2\left( (f-1)^2 - \sfrac12(1-\Omega_m) +
\Omega_k(1+\sfrac23 \Omega_m^{-1}f^2)\right), \\
T_3(x) &:= -\sfrac13 g^2\left(1-\sfrac{4}{3}\Omega_m^{-1}f^2 \right), \\
T_4(x) &:= \sfrac43 g^2\left(1+ \sfrac{2}{3}\Omega_m^{-1}f^2\right).
\end{align}
\end{subequations}
Substituting~\eqref{S_spec} in~\eqref{bbS_gen} leads to
\begin{subequations}
\be \label{bbS_spec1}  {\mathbb S}(x,x^i)= {\hat T}_1\zeta^2+
{\hat T}_2{\cal D}(\zeta) + m^{-2}\left[{\hat T}_3({\bf D}\zeta)^2 +
{\hat T}_4{\bf D}^2 {\cal D}(\zeta)\right],   \ee
where
\be {\hat T}_A(x):=\int^x_0T_A({\bar x})d{\bar x},\quad A=1,\dots,4.\label{T_hat} \ee
\end{subequations}
 Next, on substituting~\eqref{bbS_spec1} in~\eqref{Psi2_gen} we obtain
\begin{subequations}
\be  {}^{(2)}\!\Psi_{\mathrm p}=g\left( { B}_1(x)\zeta ^2+ {B}_2(x) {\cal
D}(\zeta)+ m^{-2}[{ B}_3(x)({\bf D}\zeta)^2 + { B}_4(x){\bf D}^2 {\cal
D}(\zeta)] \right),  \label{Psi2_spec1}   \ee
where
\be  \label{B_A}  {B}_A(x):= \frac{\cal H}{x^2 g}\int_0^x \frac{{\bar x}\,\Omega_m}{\cal H}  {\hat T}_A(\bar x)d{\bar x},  \quad A=1,\dots,4.   \ee
\end{subequations}
At this stage it is convenient to rescale the coefficients and
write equations~\eqref{Psi2_spec1} and~\eqref{bbS_spec1} in the form
\begin{subequations} \label{psi2.bbS}
\begin{align}
{}^{(2)}\!\Psi_{\mathrm p}&= g\!\left({\mathcal B}_1\zeta ^2+ {\mathcal B}_2 {\cal
D}(\zeta)+\sfrac23 m^{-2}xg\left[{\mathcal B}_3({\bf D}\zeta)^2 + {\mathcal
B}_4{\bf D}^2 {\cal D}(\zeta)\right]\right),     \label{Psi2_spec2}  \\
{\mathbb S}&= {\mathcal T}_1\zeta^2+
{\mathcal T}_2{\cal D}(\zeta) + \sfrac23m^{-2}xg\left({\mathcal T}_3({\bf D}\zeta)^2 +
{\mathcal T}_4{\bf D}^2 {\cal D}(\zeta)\right), \label{bbS_spec2}
\end{align}
\end{subequations}
where
\begin{subequations} \label{cal_BT}
\begin{align}
{\mathcal B}_{1,2}:&=B_{1,2}, \qquad  {\cal B}_{3,4}:=(\sfrac23xg)^{-1}B_{3,4}. \label{cal_B} \\
{\mathcal T}_{1,2}:&={\hat T}_{1,2}, \qquad  {\mathcal T}_{3,4}:=(\sfrac23xg)^{-1}{\hat T}_{3,4}.  \label{cal_T}
\end{align}
\end{subequations}
We can now calculate ${}^{(2)}\!{\bf v}_{\mathrm p}$ and
${}^{(2)}\!{\bdelta}_{\mathrm p}$ by substituting~\eqref{psi2.bbS}
into~\eqref{delta2.v2} and using the first order solution~\eqref{p1} to
calculate the source terms~\eqref{source} and~\eqref{source_dv}.\footnote
{Use~\eqref{omega_def} to express $K$ in terms of $\Omega_k$ and then use~\eqref{omega_m}
to eliminate ${\cal H}^2$.
In order to simplify the term in~\eqref{source_dv2} containing the mode extraction operator ${\cal S}^i$ it is necessary
to use the identity~\eqref{D4}.} The results are as follows.
For ${}^{(2)}\!{\bf v}_{\mathrm p}$ we obtain:
\begin{subequations} \label{vp2}
\be  {\cal H}{}^{(2)}\!{\bf v}_{\mathrm p}= { V}_1\zeta ^2+ {V}_2 {\cal
D}(\zeta)+ \sfrac23m^{-2}gx[{ V}_3({\bf D}\zeta)^2 + { V}_4{\bf D}^2 {\cal
D}(\zeta)], \ee
where
\be V_A(x):=-\sfrac23{\mathcal T}_A +(1+\sfrac23\Omega_k \Omega_m^{-1})g{\mathcal B}_A+\sfrac23 \Omega_m^{-1} g\,{\mathcal V}_A ,  \ee
with
\begin{align}
{\mathcal V}_1&=g\left(1 -2\Omega_m^{-1}f(f+\Omega_m+\Omega_k)\right),  \\
{\mathcal V}_2&= 4g\left(1+ \sfrac23 \Omega_m^{-1} f(f-\Omega_k)\right), \\
{\mathcal V}_3&= -\sfrac13 f, \qquad {\mathcal V}_4=\sfrac43 f.
\end{align}
\end{subequations}
For the density perturbation ${}^{(2)}\!{\bdelta}_{\mathrm p}$  we
obtain~\eqref{delta_2} with $_{\bullet}$ replaced by $_{\mathrm p}$, with the coefficients
$A_{i,{\mathrm p}}$ having the following form:
\begin{subequations} \label{A_poisson1}
\begin{align}
A_{1,{\mathrm p}}&=-2{\mathcal T}_1 + 3g{\mathcal B}_1 +2\Omega_m^{-1}g^2\left((1- f)^2-4\Omega_k\right), \label{A_p1}   \\
A_{2,{\mathrm p}}&= -2{\mathcal T}_2 + 3g{\mathcal B}_2 +8\Omega_m^{-1}g^2(1+\sfrac{2}{3}\Omega_m^{-1} f^2), \label{A_p2}   \\
A_{3,{\mathrm p}}&= -2{\mathcal T}_3 + 3g{\mathcal B}_3 - g\left(5+\sfrac{4}{3} \Omega_m^{-1}f^2\right),\\
A_{4,{\mathrm p}}&= -2{\mathcal T}_4 + 3g{\mathcal B}_4 + {\mathcal B}_2,  \\
A_{5,{\mathrm p}}&= {\mathcal B}_1 +4g.
\end{align}
\end{subequations}
It should be noted that these coefficients give a \emph{particular solution} for
${}^{(2)}\!{\bdelta}_{\mathrm p}$ that corresponds to the particular
solution~\eqref{Psi2.bbS} for ${}^{(2)}\!{\Psi}_{\mathrm p}$. We will give
the general solution in section~\ref{sec:initial}. The Einstein-de Sitter
background arises as a special case ($\Omega_m=1, \Omega_{\Lambda}=0,
\Omega_k=0$). It follows from~\eqref{fg}, in conjunction with~\eqref{omega_m}
and~\eqref{calH}, that $g=\sfrac35, f=1$. The definition~\eqref{cal_BT} then
yields ${\cal B}_1={\cal B}_2=0, {\cal B}_3=\sfrac{1}{21}, {\cal
B}_4=\sfrac{20}{21}$ and ${\cal T}_1={\cal T}_2=0, {\cal T}_3=\sfrac{1}{10},
{\cal T}_4=2$. The expression for ${}^{(2)}\!{\bdelta}_{\mathrm p}$ reduces
to equation (44) in UW.

\subsection{The uniform curvature and total matter gauges}

The second order density perturbations ${}^{(2)}\!\bdelta_{\mathrm c}$ and ${}^{(2)}\!{\bdelta}_{\mathrm v}$
 in {the uniform curvature and total matter gauges
are related to ${}^{(2)}\!\bdelta_{\mathrm p}$ according to
equations~\eqref{ptoc2} and~\eqref{ptov2} in appendix~\ref{sec:transf}, which
we for the readers convenience repeat here:
\begin{subequations}   \label{transf}
\begin{align}
{}^{(2)}\!{\bdelta}_{\mathrm c} &= {}^{(2)}\!{\bdelta}_{\mathrm p} - 3{}^{(2)}\!{\Psi}_{\mathrm p} +
{\mathbb S}_\delta[{}^{(1)}\!{\bf Z}_{{\mathrm c},{\mathrm p}}], \\
{}^{(2)}\!{\bdelta}_{\mathrm v} &= {}^{(2)}\!{\bdelta}_{\mathrm p} - 3{\cal H}{}^{(2)}\!{\bf v}_{\mathrm p} +
{\mathbb S}_\delta[{}^{(1)}\!{\bf Z}_{{\mathrm v},{\mathrm p}}].
\end{align}
\end{subequations}
The source terms ${\mathbb S}_\delta[{}^{(1)}\!{\bf Z}_{{\mathrm c},{\mathrm
p}}]$ and ${\mathbb S}_\delta[{}^{(1)}\!{\bf Z}_{{\mathrm v},{\mathrm p}}]$
are given by~\eqref{source_cp} and~\eqref{source_vp} in terms of
${}^{(1)}\!\Psi_{\mathrm p}, {}^{(1)}\!{\bf v}_{\mathrm p}$ and
${}^{(1)}\!{\bdelta}_{\mathrm p}$, which in the case of zero decaying mode
are given by~\eqref{p1}. It follows that the source terms are a linear
combination of the expressions for $\zeta$ in~\eqref{quad_zeta} of weights
zero and two in ${\bf D}_i$, as are ${}^{(2)}\!{\Psi}_{\mathrm p}$ and
${}^{(2)}\!{\bf v}_{\mathrm p}$ which are given by~\eqref{Psi2_spec2}
and~\eqref{vp2}. Equations~\eqref{transf} thus imply that
\emph{${}^{(2)}\!{\bdelta}_{\mathrm c}$ and ${}^{(2)}\!{\bdelta}_{\mathrm v}$
are of the canonical form~\eqref{delta_2}}, with the Newtonian terms being
unaffected by the change of gauge. The coefficients $A_{i,{\mathrm c}}$ and
$A_{i,{\mathrm v}}$ are obtained by appropriately collecting terms on the
right side of equations~\eqref{transf}, leading to
\begin{subequations} \label{A_curvature1}
\begin{align}
A_{1,{\mathrm c}}&=-2{\mathcal T}_1 + 2\Omega_m^{-1} g^2\left( 1+4f+f^2+2\Omega_k+\sfrac32\Omega_m\right), \label{A_c1}  \\
A_{2,{\mathrm c}}&= -2{\mathcal T}_2 +8\Omega_m^{-1}g^2(1+\sfrac{2}{3}\Omega_m^{-1} f^2), \label{A_c2}  \\
A_{3,{\mathrm c}}&= -2{\mathcal T}_3 +
g\left(1- 2f- \sfrac32\Omega_m- \sfrac43 \Omega_m^{-1}f^2\right),\\
A_{4,{\mathrm c}}&= -2{\mathcal T}_4+  {\mathcal B}_2+ \sfrac32 g\,\Omega_m,  \\
A_{5,{\mathrm c}}&= {\mathcal B}_1 +g(1+f),
\end{align}
\end{subequations}
and
\begin{subequations} \label{A_comoving1}
\begin{align}
A_{1,{\mathrm v}}&=2\kappa\, xg\!\left[ {\mathcal B}_1 -\sfrac23 \Omega_m^{-1}\,g(f^2- 3f-6\Omega_m)\right],\\
A_{2,{\mathrm v}}&=2\kappa\, xg\!\left[{\mathcal B}_2 - \sfrac83 \,\Omega_m^{-1}\,gf\right], \\
A_{3,{\mathrm v}}&=2\kappa\, xg\,{\mathcal B}_3 -\sfrac53 \Omega_m^{-1} g\left(2f+3\Omega_m \right),\\
A_{4,{\mathrm v}}&= 2\kappa\, xg\,{\mathcal B}_4 + {\mathcal B}_2 - \sfrac83 \Omega_m^{-1} gf , \\
A_{5,{\mathrm v}}&= {\mathcal B}_1 -\sfrac23\Omega_m^{-1}g (f^2 -3f-6\Omega_m).
\end{align}
\end{subequations}
Here we have used the fact that
\begin{subequations} \label{kappa}
\be \Omega_k \Omega_m^{-1}=-\kappa x. \ee
where the constant $\kappa$ is given by
\be \kappa: =\frac{K}{m^2},  \ee
\end{subequations}
%

\subsection{The synchronous-comoving gauge}

The second order density perturbation ${}^{(2)}\!\bdelta_{\mathrm s}$
 in \emph{the synchronous-comoving gauge}
is related to ${}^{(2)}\!\bdelta_{\mathrm v}$ according to
equation~\eqref{bdelta_2s} in appendix~\ref{sec:transf}, which we repeat here:
\be  {}^{(2)}\!{\bdelta}_\mathrm{s} = {}^{(2)}\!{\bdelta}_\mathrm{v} -
\sfrac43 xm^{-2}({\bf D}^i{\bdelta}_\mathrm{v}) ({\bf D}_i \Psi_{\mathrm p}).
\ee
On evaluating the source term using~\eqref{psi_p},~\eqref{bdelta_1} and
the identity~\eqref{D4}  in appendix~\ref{sec:opdef}, and noting that
${}^{(2)}\!\bdelta_{\mathrm v}$ has the general form~\eqref{delta_2}, we
obtain:
\begin{subequations} \label{synchr}
\be \begin{split} {}^{(2)}\!{\bdelta}_{\mathrm s} &= A_{1,{\mathrm v}} \zeta^2+ A_{2,{\mathrm v}}{\cal D}(\zeta) \\
&\quad  + \sfrac23 m^{-2}xg\left[A_{3,{\mathrm s}}({\bf D}\zeta)^2+
A_{4,{\mathrm s}}{\bf D}^2{\cal D}(\zeta)+A_{5,{\mathrm v}}{\bf D}^2 \zeta^2 \right]  \\
&\quad  +\sfrac49 m^{-4}x^2g^2\left[({\mathcal B}_3+\sfrac13){\bf D}^2({\bf D}\zeta)^2  +
({\mathcal B}_4-\sfrac43){\bf D}^4{\cal D}(\zeta) + 2({\bf D}^2\zeta)^2 \right],
\end{split}  \label{delta_2s}  \ee
where
\begin{align}  \label{A34_s}
A_{3,{\mathrm s}} &= A_{3,{\mathrm v}}-4\kappa \,xg,  \\
A_{4,{\mathrm s}} &= A_{4,{\mathrm v}}-\sfrac83\kappa\, xg,
\end{align}
\end{subequations}
and the $A_{i,{\mathrm v}}$ coefficients are given by~\eqref{A_comoving1}.
Note the appearance of the additional quadratic differential expression
$({\bf D}^2\zeta)^2$ in the Newtonian part. Equations~\eqref{bdelta_1}
and~\eqref{synchr} highlight important similarities and differences between
the total matter gauge and the synchronous-comoving gauge for dust
perturbations, namely, at first order the fractional density perturbations
are the same while at second order only the superhorizon parts coincide.

\subsection{A second order conserved quantity ${}^{(2)}\!{\bdelta}_{\mathrm c}$}

We have shown earlier that the super-horizon term in
${}^{(1)}\!{\bdelta}_{\mathrm c}$ is independent of time. This reflects the
fact that $\zeta_{\rho}:=-\sfrac13{}^{(1)}\!{\bdelta}_{\mathrm c}$ satisfies
a `conservation law' that reduces to $\partial_{\eta} \zeta_{\rho}=0$ when
spatial derivative terms can be neglected. At second order, we conjecture
that ${}^{(2)}\!\zeta_{\rho}:=-\sfrac13{}^{(2)}\!{\bdelta}_{\mathrm c}$ has a
similar property, in other words that the super-horizon terms in
${}^{(2)}\!{\bdelta}_{\mathrm c}$, namely ${A}_{1,{\mathrm c}}$ and
${A}_{2,{\mathrm c}}$, as given by~\eqref{A_c1} and~\eqref{A_c2}, are
independent of time. We can confirm this by showing directly that
\be \partial_x{A}_{1,{\mathrm c}}=0, \qquad  \partial_x{A}_{2,{\mathrm c}}=0. \ee
This calculation uses the fact that $\partial_x {\mathcal T}_{1,2}=T_{1,2}$,
the constraint~\eqref{fg2} and the following derivatives:\footnote
{Equation~\eqref{fg} gives~\eqref{evol.1}, apply $\partial_x$ to~\eqref{omega_m} and
use~\eqref{defq} to get~\eqref{evol.2}, and finally apply $\partial_x$ to~\eqref{fg2}
to get~\eqref{evol.3}.}
\begin{subequations}
\begin{align}
x\partial_x \Omega_m &= (2q-1)\Omega_m,  \label{evol.2} \\
x\partial_x g &=g(f-1), \label{evol.1} \\
x\partial_x f &=(1+f)(q-2-f) +2f+3 -\Omega_k, \label{evol.3}
\end{align}
\end{subequations}
where the deceleration parameter $q$ is defined by\footnote{This is
equivalent to $q=-\frac{\ddot {a}a}{(\dot{a})^2}. $ }
\begin{subequations}
\be x\partial_x {\cal H}=-q{\cal H}. \label{defq} \ee
It follows from~\eqref{calA} that for dust
\be q=\sfrac32 \Omega_m+\Omega_k-1. \label{q_dust} \ee
\end{subequations}
We can then determine the constant values of ${A}_{1,{\mathrm c}}$
and ${A}_{2,{\mathrm c}}$ by evaluating the limit of the
expressions~\eqref{A_c1} and~\eqref{A_c2} as $x\rightarrow 0$, leading to
\be {A}_{1,{\mathrm c}}=\sfrac{27}{5}, \qquad {A}_{2,{\mathrm c}}=\sfrac{24}{5}. \label{simple_c}  \ee
When these values are substituted into~\eqref{A_c1} and~\eqref{A_c2} we obtain
\begin{subequations} \label{T12}
\begin{align}
{\mathcal T}_1 &= \Omega_m^{-1} g^2\left( 1+4f+f^2+2\Omega_k+\sfrac32\Omega_m\right) -\sfrac{27}{10}, \\
{\mathcal T}_2 &=4\Omega_m^{-1}g^2(1+\sfrac{2}{3}\Omega_m^{-1} f^2) -\sfrac{12}{5}.
\end{align}
\end{subequations}

We have thus shown that ${}^{(1)}\!{\bdelta}_{\mathrm c}$ and ${}^{(2)}\!{\bdelta}_{\mathrm c}$
 are conserved quantities in the sense that the
super-horizon part is constant in time. These conserved quantities are in fact closely related
to the quantities introduced by Malik and Wands:\footnote
{See~\cite{malwan09}, equations (7.61) and (7.71). The
subscript $_\rho$ stands for the uniform density gauge, and the subscript
$_\mathrm {mw}$ stands for Malik-Wands. The gauge invariant ${}^{(2)}\!\zeta_{\mathrm{mw}}$
was first defined in~\cite{malwan04} (see equation (4.18). We note that Langlois and Vernizzi~\cite{lanver05}
derived an analogous conserved quantity at second order using the $1+3$ approach to
perturbations (see equation (50)).}
\be {}^{(1)}\!\zeta_{\mathrm{mw}}:=-{}^{(1)}\!\Psi_{\rho}, \quad
{}^{(2)}\!\zeta_{\mathrm{mw}}:=-{}^{(2)}\!\Psi_{\rho},  \label{zeta_mw} \ee
which we shall use in the following section. At first order we have the
simple relation ${}^{(1)}\!\Psi_{\rho}=-\sfrac13{}^{(1)}\!{\bdelta}_{\mathrm c}$,
as follows from~\eqref{bdelta_1} and~\eqref{psi_rho1}. At second order these
conserved quantities are related through their super-horizon terms:
\be \left({}^{(2)}\!\Psi_{\rho} - {}^{(1)}\!\Psi_{\rho}^2\right)|_{\text{super-horizon}}=
-\sfrac13 {}^{(2)}\!{\bdelta}_{\mathrm c}|_{\text{super-horizon}},  \ee
as follows from~\eqref{simple_c},~\eqref{bdel2_gen} and~\eqref{psi_rho2}.

\subsection{Initial conditions} \label{sec:initial}

The solution~\eqref{Psi2.bbS} for ${}^{(2)}\!\Psi_{\mathrm p}$ is a
particular solution that satisfies $\lim_{x\rightarrow
0}{}^{(2)}\!\Psi_{\mathrm p}=0$. The general solution for
${}^{(2)}\!\Psi_{\mathrm p}$ for a zero decaying scalar mode at the linear
level is given by
\be {}^{(2)}\!\Psi_{\mathrm p}|_{gen} ={}^{(2)}\!\Psi_{\mathrm p} +
C(x^i)g(x), \label{psi2_gen}  \ee
where $C(x^i)$ is an arbitrary function. Note that the second term on the
right side of~\eqref{psi2_gen} is the general solution of the homogeneous
equation for ${}^{(2)}\!\Psi_{\mathrm p}$ (see UW, equation (37)). The
corresponding general expression for ${}^{(2)}\!{\bdelta}_{\bullet}$ for the
Poisson, the uniform curvature, the total matter and the synchronous-comoving gauges
is given by
\be  {}^{(2)}\!{\bdelta}_{\bullet}|_{gen}={}^{(2)}\!{\bdelta}_{\bullet} + A_{\bullet}C +
\sfrac23xgm^{-2}{\bf D}^2C,   \label{bdel2_gen}  \ee
where $A_{\bullet}$ is the coefficient in the expression~\eqref{bdelta_1} for
${}^{(1)}\!{\bdelta}_{\bullet}$. Note that the extra terms on the right side
of~\eqref{bdel2_gen} take the same form as the first order density
perturbation, but with $\zeta(x^i)$ replaced by the arbitrary function
$C(x^i)$.

In applications the arbitrary function $C(x^i)$ is usually determined by
using the metric and matter perturbations at the end of inflation as initial
conditions. Various theories of inflation predict that these perturbations
will not be purely Gaussian \emph{i.e.} there will be a certain level of
primordial non-Gaussianity. It is convenient to use the first and second
order conserved quantities given by~\eqref{zeta_mw} to parameterize this
primordial non-Gaussianity on super-horizon
scales. Specifically, it is assumed that
\be
{}^{(2)}\!\zeta_{\mathrm{mw}}=2a_{\mathrm{nl}}\left({}^{(1)}\!\zeta_{\mathrm{mw}}\right)^2,
\label{png} \ee
where $a_{\mathrm{nl}}$ is a parameter that depends on the physics of the
model of inflation.\footnote{See for example~\cite{baretal04a}, page 4,
and~\cite{baretal10}, equation (9), and references given in these papers.
When making comparisons with CMB observations it is customary to use a
non-linearity parameter $f_{\mathrm{nl}}$, which takes into account that the
nonlinear gravitational dynamics after inflation contributes to the
non-Gaussianity. This parameter has the form $f_{\mathrm{nl}}=\sfrac53(
a_{\mathrm{nl}} -1)+\dots,$ where $+\dots$ refers to terms that describe the
effect of the post-inflation nonlinear gravitational dynamics on the
primordial non-Gaussianity. See for example~\cite{baretal04a}, equation
(9),~\cite{baretal10}, equation (31) and~\cite{baretal04b}, section 8.4.2.}
It has been shown that primordial non-Gaussianity in the CMB temperature
anisotropy at second order is represented by the quantity $
{}^{(2)}\!\zeta_\mathrm {mw}-2 {}^{(1)}\!\zeta_\mathrm {mw}^2$
(\cite{baretal04a}, equation (8)). It follows that the absence of primordial
non-Gaussianity corresponds to $a_{\mathrm{nl}}=1$.

It follows from equations~\eqref{Psi_rho} in appendix~\ref{sec:transf} in conjunction
with~\eqref{psi_p} and~\eqref{bdelta_1} that the gauge invariants ${}^{(1)}\!{\Psi}_{\rho}$
and ${}^{(2)}\!{\Psi}_{\rho}$ in~\eqref{zeta_mw} are given by
\begin{subequations}
\begin{align}
{}^{(1)}\!{\Psi}_{\rho} &= \zeta - \sfrac29 xg\,m^{-2}\,{\bf D}^2 \zeta,
\label{psi_rho1} \\
{}^{(2)}\!{\Psi}_{\rho} &= -\sfrac15[4\zeta^2 +8 {\cal D}(\zeta)] +
C(x^i) + ({\bf D}_i\, \, \text{terms up to order}\, \,6).  \label{psi_rho2}
\end{align}
\end{subequations}
These equations and the restriction~\eqref{png} determine the arbitrary
function $C(x^i)$ in terms of $\zeta$ and ${\cal D}(\zeta)$. The resulting
function is denoted by $C_{\mathrm{nl}}$:
\be C_{\mathrm{nl}}(x^i)=\sfrac45 [(1-\sfrac52 a_{\mathrm{nl}})\zeta^2 +2{\cal D}(\zeta)]. \label{C_png} \ee
We will denote the density perturbation ${}^{(2)}\!{\bdelta}_{\bullet}$
corresponding to this choice of initial condition, which is is determined by
substituting the expression~\eqref{C_png} into~\eqref{bdel2_gen}, by
${}^{(2)}\!{\bdelta}_{\bullet}|_{\mathrm{nl}}$:
\be  {}^{(2)}\!{\bdelta}_{\bullet}|_{\mathrm{nl}}={}^{(2)}\!{\bdelta}_{\bullet} + A_{\bullet}C_{\mathrm{nl}} +
\sfrac23xgm^{-2}{\bf D}^2C_{\mathrm{nl}}.   \label{bdel2_ng}  \ee
It follows from~\eqref{delta_2},~\eqref{bdel2_ng}
and~\eqref{C_png} that the coefficients $A_{i,{\bullet}}|_{\mathrm{nl}}$ are given by
\begin{subequations} \label{init}
\begin{align}
A_{1,{\bullet}}|_{\mathrm{nl}}&= A_{1,{\bullet}} +\sfrac45(1-\sfrac52 a_{\mathrm{nl}})A_{\bullet}, \\
A_{2,{\bullet}}|_{\mathrm{nl}}&=  A_{2,{\bullet}} +\sfrac85 A_{\bullet}, \\
A_{3,{\bullet}}|_{\mathrm{nl}}&=  A_{3,{\bullet}}, \\
A_{4,{\bullet}}|_{\mathrm{nl}}&=  A_{4,{\bullet}} + \sfrac{8}{5},\\
A_{5,{\bullet}}|_{\mathrm{nl}}&=   A_{5,{\bullet}} +\sfrac{4}{5}(1-\sfrac52 a_{\mathrm{nl}}),
\end{align}
\end{subequations}
where $A_{\bullet}$ is given by~\eqref{bdelta_1}.

\section{The specialization to a flat background}\label{sec:K0}

In the previous section we showed that the time dependence of $
{}^{(2)}\!{\bdelta}_{\bullet}$ is determined by the linear perturbation function
$g(x)$ and the background functions $\Omega_m$ and ${\cal H}$, either
algebraically, or as the integrals ${\mathcal B}_A$ given by~\eqref{B_A}
and~\eqref{cal_B}, and ${\mathcal T}_A$ given by~\eqref{T_hat}
and~\eqref{cal_T}, with $A=1,\dots ,4$. Subsequently, we showed that
${\mathcal T}_1$ and ${\mathcal T}_2$ could be written algebraically, as
in~\eqref{T12}. In this section we show that if the spatial curvature is
zero, a significant simplification occurs: \emph{only one integral function
is required}.

\subsection{The flatness conditions}

We here show that if the background is flat then ${\mathcal T}_{3,4}$ and
${\mathcal B}_{1,2}$ are algebraic expressions in $g$ and $\Omega_m$, and in
addition ${\mathcal B}_3 +{\mathcal B}_4=1$. For convenience we define:
\begin{subequations}  \label{T34B12}
\begin{align}
{\bf T}_3&:={\mathcal T}_3 - \sfrac12 g +\sfrac34\Omega_m g^{-1}(1-g)^2,  \label{T3} \\
{\bf T}_4&:={\mathcal T}_4 -3 +2g - \sfrac34\Omega_m g^{-1}(1-g)^2,\\
{\bf B}_1&:= {\mathcal B}_1  - \sfrac15 +g - \sfrac32\Omega_m g^{-1}(1-g)^2,\\
{\bf B}_2&:= {\mathcal B}_2 -\sfrac{12}{5} +4g, \\
{\bf B}_{3+4}&:=xg({\mathcal B}_3 +{\mathcal B}_4-1).
\end{align}
\end{subequations}
Then the result can be expressed as follows: if $\Omega_k=0$ then
${\bf T}_{3,4}=0, \,{\bf B}_{1,2}=0$, and ${\bf B}_{3+4}=0$.

These results can be proved by differentiation, as follows. First we show
that if $K=0$ then $\partial_x(xg{\bf T}_{3,4})=0.$ This calculation requires
$\partial_x(xg{\mathcal T}_{3,4})=\sfrac32T_{3,4}$, as follows
from~\eqref{T_hat} and~\eqref{cal_T}, and also equations~\eqref{evol.1}
and~\eqref{evol.2}. It follows that $xg{\bf T}_{3,4}=C_{3,4}$, a constant.
Since $T_{3,4}$ is bounded as $x\rightarrow 0$ we conclude that $C_{3,4}=0$,
which gives the desired result.

Second we show that if $K=0$ then the quantities ${\bf B}_1, {\bf B}_2$ and
${\bf B}_{3+4}$ satisfy
\be x\partial_x {\bf B}_{\bullet}=-\sfrac32\Omega_m g^{-1}{\bf B}_{\bullet},
\qquad \lim_{x\rightarrow 0}{\bf B}_{\bullet}=0,
\label{B_bullet}   \ee
Since $g>0$ it follows from~\eqref{B_bullet} that $ ({\bf B}_{\bullet})^2$ is
monotone decreasing or identically zero. The limit condition then implies
that $ {\bf B}_{\bullet}\equiv 0$, which gives the desired relations ${\bf
B}_{1,2}=0$, and ${\bf B}_{3+4}=0$. This calculation requires
\be \partial_x(x^2g{\cal H}^{-1}{B}_A)= x\Omega_m{\cal H}^{-1}{\hat T}_A,
\qquad
\partial_x(x^2g{\cal H}^{-1})= \sfrac32 x\Omega_m{\cal H}^{-1}. \label{partialB_A} \ee
The first of these follows from~\eqref{B_A}, together with the
definitions~\eqref{cal_B} and~\eqref{cal_T}. The second follows
from~\eqref{evol.2},~\eqref{defq} and~\eqref{fg2}. Note that the
constraint~\eqref{fg2} can be written in the form $2g(f+q+1)=3\Omega_m$,
using~\eqref{q_dust}.

\subsection{Alternate expressions for ${}^{(2)}\!{\bdelta}_{\bullet}$}

We now cast the expressions for ${}^{(2)}\!{\bdelta}_{\bullet}$ into a form in
which the role played by the spatial curvature becomes clear. We
use~\eqref{T12} to eliminate ${\mathcal T}_{1,2}$ in the expressions for
${}^{(2)}\!{\bdelta}_{\bullet}$ and use~\eqref{T34B12} to express ${\mathcal
T}_{3,4}$ and ${\mathcal B}_{1,2}$ in terms of ${\bf T}_{3,4}$ and ${\bf
B}_{1,2}$. We also use the constraint~\eqref{fg2} to eliminate $f$ in favour
of $g$, and use~\eqref{init} to introduce the non-Gaussianity initial
condition. The coefficients $A_{i,{\bullet}}$ in
equations~\eqref{A_poisson1},~\eqref{A_curvature1} and~\eqref{A_comoving1}
assume the form
\begin{subequations}  \label{A_poisson}
\begin{align}
A_{1,{\mathrm p}} &= 3(1-g)\left( 1 +2a_{\mathrm{nl}} - 4g +\sfrac32\Omega_m(1-g) \right) +3g\,{\bf B}_1, \\
A_{2,{\mathrm p}} &= 12g(1-g)  +3g\,{\bf B}_2,  \\
A_{3,{\mathrm p}} &=3g({\mathcal B}_3-2)-\sfrac32\Omega_m\, g^{-1}(1-g)^2
+4 \Omega_k(1-g+\sfrac13\kappa x g) -2{\bf T}_3, \\
A_{4,{\mathrm p}} &= -2+ 3g{\mathcal B}_4 -\sfrac32\Omega_m\,g^{-1}(1-g)^2 -2{\bf T}_4 +{\bf B}_2,, \\
A_{5,{\mathrm p}} &=1-2a_{\mathrm{nl}} + 3g+ \sfrac32\Omega_m\,g^{-1}(1-g)^2+{\bf B}_1,
\end{align}
\end{subequations}
\begin{subequations}  \label{A_curvature}
\begin{align}
A_{1,{\mathrm c}} &=3(1+2 a_{\mathrm{nl}}), \\
A_{2,{\mathrm c}}&=0, \\
A_{3,{\mathrm c}} &=- \sfrac32 \Omega_m\,g^{-1} +2\Omega_k\,(2-g+\sfrac23\kappa x g)-2{\bf T}_3, \\
A_{4,{\mathrm c}} &= - 2 -\sfrac32 \Omega_m\,g^{-1}(1-2g) -2{\bf T}_4 +{\bf B}_2, \\
A_{5,{\mathrm c}} &= 1-2 a_{\mathrm{nl}} +\sfrac32 \Omega_m\,g^{-1}(1-g) -\Omega_k\,g +{\bf B}_1,
\end{align}
\end{subequations}
\begin{subequations} \label{A_comoving}
\begin{align}
A_{1,{\mathrm v}}&=2\kappa\, xg\!\left[ {\bf B}_1 +\sfrac{16}{5}+ 2\kappa\, xg (1+\sfrac13\Omega_k) +2\Omega_k (1-g) )\right],\\
A_{2,{\mathrm v}}&=2\kappa\, xg\!\left[{\bf B}_2-\sfrac85 - \sfrac83 \kappa\,xg\right], \\
A_{3,{\mathrm v}}&=2\kappa\, xg\,({\mathcal B}_3-\sfrac53) -5,\\
A_{4,{\mathrm v}}&= 2\kappa\, xg\,({\mathcal B}_4 -\sfrac43) + {\bf B}_2, \\
A_{5,{\mathrm v}}&=2(2-a_{nl}) + 2\kappa\, xg (1+\sfrac13\Omega_k) +2\Omega_k (1-g) +{\bf B}_1,
\end{align}
\end{subequations}
where the constant $\kappa$ is defined by~\eqref{kappa}. For the synchronous-comoving
gauge it follows from~\eqref{synchr} that
\be (A_1, A_2, A_3, A_4, A_5)_{\mathrm s}  =(A_1, A_2, A_3, A_4 , A_5)_{\mathrm v} -
4\kappa\,xg(0,0,1 ,\sfrac23, 0),  \label{A_synch} \ee
with the Newtonian part unchanged.

\subsection{Zero spatial curvature}

In the case of zero spatial curvature  we have $\Omega_k=0,\,\kappa=0,{\bf
T}_{3,4}=0,\,{\bf B}_{1,2}=0$ and ${\mathcal B}_3+{\mathcal B}_4=1.$ We write
${\mathcal B}_3={\mathcal B}$ and ${\mathcal B}_4=1-{\mathcal B}$, and can
express the scalar ${\mathcal B}$ as a standard integral involving $g,
\Omega_m$ and ${\cal H}$:
\be {\mathcal B}(x)=\frac{{\cal H}(x)}{2x^3g(x)^2}\int_0^x
\frac{{\bar x}^2\,\Omega_m}{\cal H}\left( g^2-\sfrac32\Omega_m (1-g)^2  \right)d{\bar x}, \label{cal _Bfinal}  \ee
where $g, \Omega_m$ and ${\cal H}$ inside the integral are functions of
${\bar x}$. This result follows from equations~\eqref{B_A} and~\eqref{cal_BT}
with $A=3$, when one uses the expression for ${\mathcal T}_3$ given
by~\eqref{cal_T} and~\eqref{T3} with ${\bf T}_3=0$.

With these simplifications the
expressions~\eqref{A_poisson}-\eqref{A_comoving} for the coefficients in
${}^{(2)}\!{\bdelta}_{\bullet}$ for the Poisson, uniform curvature and total
matter gauges reduce to those in~\cite{uggwai13c}, where the present results
for zero curvature are summarized. The full expression is given
by~\eqref{delta_2}, with the Newtonian part given by
\be  {}^{(2)}\!{\bdelta}_{\bullet}|_{Newtonian} =
{\sfrac49m^{-4}x^2g^2\left[{\mathcal B}\,{\bf D}^2({\bf D}\zeta)^2 +
(1- {\mathcal B}){\bf D}^4{\cal D}(\zeta)  \right]}.  \label{newt_special} \ee
For the reader's convenience, we give the coefficients $A_{i,{\mathrm p}}$ in the
Poisson gauge when $K=0$, obtained by specializing~\eqref{A_poisson}:
\begin{subequations}  \label{A_poisson_spec}
\begin{align}
A_{1,{\mathrm p}} &= 3(1-g)\left( 1 +2a_{\mathrm{nl}} - 4g +\sfrac32\Omega_m(1-g) \right), \\
A_{2,{\mathrm p}} &= 12g(1-g),  \\
A_{3,{\mathrm p}} &=3g({\mathcal B}-2)-\sfrac32\Omega_m\, g^{-1}(1-g)^2, \\
A_{4,{\mathrm p}} &= -2+ 3g(1-{\mathcal B}) -\sfrac32\Omega_m\,g^{-1}(1-g)^2, \\
A_{5,{\mathrm p}} &=1-2a_{\mathrm{nl}} + 3g+ \sfrac32\Omega_m\,g^{-1}(1-g)^2.
\end{align}
\end{subequations}
We also give the full expression for the synchronous-comoving gauge:
\begin{equation} \begin{split}
{}^{(2)}\!{\bdelta}_{\mathrm s}&=
\sfrac23 m^{-2}xg\left[-5({\bf D}\zeta)^2  + 2(2-a_{\mathrm{nl}}) {\bf D}^2 \zeta^2 \right] \\
&\qquad + \sfrac49m^{-4}x^2g^2\left[({\mathcal B}+\sfrac13)\left({\bf D}^2({\bf D}\zeta)^2  -
{\bf D}^4{\cal D}(\zeta)\right) +2({\bf D}^2\zeta)^2 \right],
\end{split}
\label{deltaN_2s}
\end{equation}
as follows from~\eqref{A_comoving} and~\eqref{A_synch}.

\subsection{Relation with the literature}

In the course of doing the research reported in this paper we made a detailed
comparison of our expressions for ${}^{(2)}\!\bdelta_{\bullet}$ with those in
the literature, which deal solely with the case where \emph{the background
spatial curvature is zero}. In addition the expressions for
${}^{(2)}\!\bdelta_{\bullet}$ with $\Lambda>0$ are \emph{restricted to the
synchronous-comoving and Poisson gauges.} In~\cite{uggwai13c} we gave a brief overview
of the results in the literature. In this section we describe in detail the
relation between our results and the papers in the literature, focussing in
particular on the work of Tomita~\cite{tom05} and Bartolo and
collaborators~\cite{baretal10}.

Comparing the different results is not straightforward since there are many
different ways of representing the spatial dependence in the expression for
${}^{(2)}\!{\bdelta}_{\bullet}$, which involves an arbitrary spatial function
and the spatial differential operator ${\bf D}_i$. We thus begin by
describing the various quadratic differential expressions that have been used
in the literature and showing how they are related to our canonical
set~\eqref{quad_zeta}.

\subsubsection{Spatial quadratic differential expressions}

In discussing our canonical set of quadratic differential
expressions~\eqref{quad_zeta} we note that the operator ${\cal D}(A)$, as
defined by~\eqref{DA} and~\eqref{modeextractop}, plays a key role. We begin
with the zero order derivative expressions $\zeta^2$ and ${\cal D}(\zeta)$
that determine the super-horizon terms in ${}^{(2)}\!{\bdelta}_{\bullet}$.
Two of the three second order derivative expressions that determine the
post-Newtonian terms are obtained by acting with ${\bf D}^2$ on the zero
order expressions, while the third, $({\bf D}\zeta)^2\equiv {\bf D}^i
\zeta{\bf D}_i \zeta$, is a new expression. Finally, the two fourth order
derivative expressions that determine the Newtonian terms are obtained by
acting with ${\bf D}^2$ on two of the second order expressions. Before
continuing we mention that the appearance of ${\cal D}(\zeta)$ in the second
order density perturbation has its origin in the quadratic source term in the
evolution equation for the second order Bardeen potential
$^{(2)}\!\Psi_{\mathrm p}$ (see equations (61b) and (61f)
in~\cite{uggwai13b}), through the use of the mode extraction operator
${\mathcal S}^{ij}$, as defined by~\eqref{modeextractop}.

We now list the various other spatial quadratic differential expressions that
have appeared in the literature:
\begin{subequations}
\begin{align}
& \quad A{\bf D}^2A, \qquad\qquad {\bf D}^i {\bf D}^j({\bf D}_iA {\bf D}_jA ), \qquad
({\bf D}^i {\bf D}^j A)({\bf D}_i{\bf D}_jA ), \\
& ({\bf D}^i A)({\bf D}_i{\bf D}^2A ), \qquad {\bf D}^i ({\bf D}_iA{\bf D}^2A
), \qquad\quad  {\bf D}^i {\bf D}^j (A{\bf D}_i{\bf D}_jA ),
\end{align}
\end{subequations}
sometimes with ${\bf D}^{-2}$ acting on the left. Each of these expressions
can be written as a linear combination of our canonical set~\eqref{quad_zeta}
augmented by the terms $({\bf D}^2 A)^2$ and ${\bf D}^4 A^2$ using the
identities~\eqref{D_iden} in the Appendix. Here we use the generic symbol
$A=A(x^i)$ to denote the arbitrary spatial function. Although there is no
consensus for this function the various choices differ only by an overall
numerical factor.

A quantity closely related to our ${\mathcal D}(A)$ in~\eqref{DA} has been
defined by several authors as follows. Let
\begin{subequations}
\begin{align}
\Psi_0 &:=\sfrac12 \lambda{\bf D}^{-2}\left( {\bf D}^i{\bf D}^j A\, {\bf D}_i{\bf D}_j A-({\bf D}^2A)^2 \right),  \label{D7}  \\
\Theta_0 &:={\bf D}^{-2}\left(\Psi_0 -\sfrac13 \lambda ({\bf D}A)^2 \right), \label{D8}
\end{align}
\end{subequations}
where $\lambda$ is a numerical factor that we have introduced to accommodate different scalings.
It follows from~\eqref{D2} that
\be \Psi_0:=-\sfrac13 \lambda\left( {\bf D}^2{\mathcal D}(A)  - ({\bf D}A)^2
\right), \qquad \Theta_0=- \sfrac13 \lambda {\mathcal D}(A).  \label{D9}
\ee
This makes clear that $\Theta_0$ corresponds to our ${\mathcal D}(A)$, while
$\Psi_0$ is closely related to our $ {\bf D}^2{\mathcal D}(A)$. These
quantities were used in~\cite{matetal98} and~\cite{baretal05} with
$A=\varphi=\sfrac35\zeta$ and $\lambda=1$ (see equations (4.36) and (6.6)
in~\cite{matetal98}, and following equation (9) in~\cite{baretal05}). They
were also used in~\cite{tom05} with $A=F=-2\zeta$ and $\lambda=\frac{9}{100}$
(see equations (4.11), noting that a factor of ${\bf D}^{-2}$ is missing in
the first equation). Furthermore, Bartolo \emph{et al}~\cite{baretal10} come
close to introducing our ${\mathcal D}(A)$. They define (see their equation
(18))
\be  \alpha_0:={\bf D}^{-2} ({\bf D}\varphi_0)^2 -3 {\bf D}^{-4} {\bf D}^i{\bf D}^j \,( {\bf D}_i \varphi_0{\bf D}_j \varphi_0),   \ee
where $\varphi_0=\sfrac35\zeta$. It follows from~\eqref{D5} that
\be \alpha_0= -2{\mathcal D}(\varphi_0).  \label{D10} \ee
%

\subsubsection{Synchronous-comoving gauge with $K=0, \Lambda\geq0$}

The first expression for ${}^{(2)}\!{\bdelta}_{\mathrm s}$ with $\Lambda>0$
was given by Tomita~\cite{tom05} (see his equation (2.22)). The time
dependence in his expression is described by two functions $P(\eta)$ and
$Q(\eta)$ that satisfy second order differential equations:\footnote {Here
$\eta$ denotes conformal time. Note that $\partial_{\eta}={\cal
H}x\partial_x.$}
\be \left( \partial_{\eta}^2+2{\cal H}\partial_{\eta}\right)P = 1, \qquad
\left( \partial_{\eta}^2+2{\cal H}\partial_{\eta}\right)Q = P-\sfrac52
(\partial_{\eta}P)^2, \ee
and the spatial dependence is described by a function $F(x^i)$ and its first
and second partial derivatives, including the following quadratic
differential expressions:
\be  F{\bf D}^2 F, \qquad {\bf D}_i{\bf D}_j F\, {\bf D}^i{\bf D}^j F,
\label{tom_space1}   \ee
in addition to the ones in our canonical list~\eqref{quad_zeta}. We use the
identities~\eqref{D1} and~\eqref{D2} to relate these expressions to our
canonical expressions. To match the  density perturbations requires
\begin{subequations}  \label{tom_rel1}
\be g=1-{\cal H}\partial_{\eta}P, \qquad F(x^i)=-2\zeta(x^i),   \label{g1}
\ee
at linear order  and
\be P=\sfrac23 m^{-2}xg, \label{g2}  \ee
\be  \partial_\eta Q= \sfrac13 m^{-2} \frac{x}{\cal H}\left[21g^2 {\mathcal B}-  g(9g-2) + \sfrac{21}{2}\Omega_m(1-g)^2 \right],  \ee
\end{subequations}
at second order. With these equations it follows that Tomita's expression
(2.22) is transformed into our expression for ${}^{(2)}\!{\bdelta}_{\mathrm
s}$ given by~\eqref{deltaN_2s}, but with $a_{\mathrm{nl}}=0$.

\subsubsection{Poisson gauge with $K=0, \Lambda\geq0$}

The first expression for $ {}^{(2)}\!{\bdelta}_{\mathrm p}$ with $\Lambda>0$
was given by Tomita~\cite{tom05} (see his equation (4.16)). As with $
{}^{(2)}\!{\bdelta}_{\mathrm s}$ the time dependence is described by $P$ and
$Q$ and the spatial dependence is described by the quadratic differential
expressions in our canonical list~\eqref{quad_zeta} together with the
expressions~\eqref{tom_space1} and
\be ({\bf D}^i F){\bf D}^2{\bf D}_i F. \label{tom_space2} \ee
In particular the  combinations $\Psi_0$ and $\Theta_0$, as defined
by~\eqref{D7} and~\eqref{D8}, are used with $F=-2\zeta$ and
$\lambda=\sfrac{9}{100}$. We write these combinations in the form~\eqref{D9},
and use the identity~\eqref{D4} for the expression in~\eqref{tom_space2}.
Using equations~\eqref{tom_rel1} we can now show that Tomita's expression
(4.16), with a few minor typos corrected\footnote {In line 2 in Tomita's
equation (4.16), $-\frac{2a'}{a}PP' $ should be $-\frac{a'}{2a}PP' $, in line
3, $-\sfrac12 P'$ should be $-P'$, and in line 4, $-\sfrac12 P$ should be
$-P$. In addition the sign of $Q$, which appears in lines 1 and 4, should be
reversed.  } is transformed into our expression for
${}^{(2)}\!{\bdelta}_{\mathrm p}$ given
by~\eqref{delta_2},~\eqref{newt_special} and~\eqref{A_poisson_spec}, but with
$a_{\mathrm{nl}}=0$.

An expression for ${}^{(2)}\!{\bdelta}_{\mathrm p}$ with $\Lambda>0$ has more
recently been derived by Bartolo and collaborators~\cite{baretal10} (see
their equation (29)).  In order to make a comparison with our expression
which has the general form~\eqref{delta_2}, we first consider their linear
perturbation. The time dependence of the linear perturbation is described by
a function $g$, which we denote by $g_{\mathrm b}$ to distinguish it from our
$g$, and the spatial dependence is described by a function $\varphi(x^i)$
which is a constant multiple of our $\zeta$. Since $g_{\mathrm b}=1$ and
$g=\sfrac35$ when $\Lambda=0$ it follows by comparing their equation (11)
with our~\eqref{psi_p} that
\be g_{\mathrm b}=\sfrac53 g,  \qquad \varphi_0=\sfrac35 \zeta.   \label{g_b}
\ee

We next consider the Bardeen potential ${}^{(2)}\!\Psi_{\mathrm p}$ (equation
(20) in~\cite{baretal10}) whose time dependence is described by $g_{\mathrm
b}$ and four functions ${\mathbb B}_A, A=1,2,3,4$. Our expression for
${}^{(2)}\!\Psi_{\mathrm p}$, including the non-Gaussianity initial
condition, is given by equations~\eqref{Psi2_spec2},~\eqref{psi2_gen}
and~\eqref{C_png}. In order to match the spatial dependence terms we note
that their $\alpha_0$ is given by~\eqref{D10}. We also need to use the
identities~\eqref{D1} and~\eqref{D5}. Comparing the two expressions for
${}^{(2)}\!\Psi_{\mathrm p}$ leads to the following relation between our
$B_A$ and the quantities ${\mathbb B}_A$ in~\cite{baretal10}:
\be (B_1, B_2, B_3, B_4) =\sfrac{9}{25} g^{-1}\left({\mathbb B}_1, -2{\mathbb B}_2,\,
 m^2(\sfrac13 {\mathbb B}_3+{\mathbb B}_4), \sfrac23 m^2{\mathbb B}_3\right).  \label{mathbbB}     \ee

To establish consistency we need to show that the definition of the ${\mathbb
B}_A$ in~\cite{baretal10} (see equations (22)-(26)) translates into our
definition of the $B_A$ in~\eqref{B_A} under the
transformation~\eqref{mathbbB}. The definitions of the  ${\mathbb B}_A$
in~\cite{baretal10} can be collectively written in the form\footnote{Here we
have used~\eqref{fg2} and $m^2=\Omega_{m,0}{\cal H}_0^2$ to write the
equation ${\tilde B}_A={\cal H}_0^2(f_0 +\sfrac32\Omega_{m,0})B_A$
in~\cite{baretal10} in the form ${\tilde B}_A=\sfrac32\ m^2 g_0^{-1}B_A.$  }
\be \sfrac{9}{25} g^{-1}{\mathbb B}_A = \left(\sfrac53\, g_0\right)
\frac{\cal H} {x^2g}\int^x_0 \left(I(x)-I(\bar x)\right) g({\bar x})^2\,
{\mathbb T}_A({\bar x})d{\bar x}, \label{bbB1} \ee
where $I(x)$ is expressed in terms of $g(x)$ by~\eqref{fg}.
The functions ${\mathbb T}_A$ are related to our functions $T_A$ according to
\be m^2 (T_1, T_2, T_3, T_4) = g^2\left({\mathbb T}_1, -2{\mathbb T}_2,\, m^2(\sfrac13
{\mathbb T}_3+{\mathbb T}_4), \sfrac23 m^2{\mathbb T}_3\right),  \label{mathbbT}
\ee
where we note that our variable $f$ coincides with the $f$
in~\cite{baretal10}. On the other hand our equation~\eqref{B_A} expressing
$B_A$ in terms of $T_A$, can be written in the equivalent
form\footnote{Equation~\eqref{B_A}, together with~\eqref{T_hat}, is an
iterated integral. Use $x\Omega_m {\cal H}^2=m^2$ in the integrand, reverse
the order of integration and make use of the definition of $I(x)$.}
\be  B_A= \frac{\cal H}{x^2 g}\int_0^x
\left(I(x)-I(\bar x)\right)m^2\, T_A(\bar x)d{\bar x}, \quad I(x):=\int_0^x \frac{d{\bar x}}{{\mathcal H}({\bar x})^3} \label{bbB2}
\ee
The common structure of~\eqref{mathbbB} and~\eqref{mathbbT} ensures
that~\eqref{bbB1} translates into equation~\eqref{bbB2}, provided that
$g_0=\sfrac35$.

We are now in a position to show that the expression for
${}^{(2)}\!{\bdelta}_{\mathrm p}$ in~\cite{baretal10} can be transformed into
our expression. We first convert the quadratic differential expressions
in~\cite{baretal10} in $\varphi_0$ into our canonical
expressions~\eqref{quad_zeta} in $\zeta$
using~\eqref{D1},~\eqref{D5},~\eqref{D10} and~\eqref{g_b}. We then express
$g_{\mathrm b}$ and ${\mathbb B}_A$ in the time-dependent coefficients in
terms of $g$ and $B_A$, using~\eqref{g_b} and~\eqref{mathbbB}. It is
necessary to use~\eqref{partial_psi} in order to eliminate the derivatives
$\partial_x{\mathbb B}_A$. We find agreement except with the coefficient of
$\zeta^2$, which corresponds to the coefficient of $\varphi^2$ in equation
(29) in~\cite{baretal10}. We conclude that the term $(f-1)^2-1$ in this
coefficient should be replaced by $2(f-1)^2$. The $\varphi^2$ term then
correctly specializes to $-\sfrac{8}{3}(1 - \sfrac25
a_{\mathrm{nl}})\varphi^2$ when one restricts consideration to the
Einstein-de Sitter universe ($\Lambda=0$), in agreement with equation (8)
in~\cite{baretal05}.

\section{Concluding remarks}\label{Sec:concl}

The results in this paper fall under three headings. First, we have presented
exact expressions for the second order fractional density perturbation for
dust, a cosmological constant and spatial curvature in a simple and
physically transparent form in four popular gauges: the Poisson, the uniform
curvature, the total matter and the synchronous-comoving gauges. Our results unify and
generalize all the known results in the literature, which are confined to the
case of zero spatial curvature and, when $\Lambda>0$, to the Poisson and
synchronous-comoving gauges. Our approach has two novel features. We have introduced a
canonical way of representing the spatial dependence of the perturbations at
second order which makes clear how the choice of gauge affects the form of
the expressions. In addition we have formulated the time dependence in such a
way that the dynamics of the perturbations and the effect of spatial
curvature can be read off by inspection. In particular, in the special case
of zero spatial curvature we have shown that the time evolution simplifies
dramatically, and requires the use of only two non-elementary functions, the
so-called growth supression factor $g$ that arises at the linear level, and
one new function ${\mathcal B}$ at the second order level. We emphasize that
the assumption of zero decaying mode underlies the simple expressions for
${}^{(2)}\!\bdelta_{\bullet}$ that we have presented. This assumption is
usually made in cosmological perturbation theory, presumably on the grounds
that the decaying mode will become negligible. However, if $\Lambda>0$ the
name "decaying mode" is a misnomer since this mode, after decaying in the
matter-dominated epoch ($\Omega_m\approx1$), increases when $\Omega_\Lambda$
becomes significant and contributes to the density perturbation on an equal
footing with the growing mode in the de Sitter regime. This is made clear by
the asymptotic expressions given in UW (see equation (66a)). Into the past
the decaying mode grows without bound on approach to the initial singularity.
On the other hand, if the decaying mode is set to zero, the perturbations
remain finite into the past and one is essentially considering perturbations
in a universe with an isotropic singularity~\cite{goowai85}.

Second, we have made a detailed comparison  of our results with the known
expressions for ${}^{(2)}\!\bdelta$ in different gauges when the background
spatial curvature is zero. Our canonical representation of the spatial
dependence has enabled us to unify seemingly disparate results, while at the
same time revealing a number of errors in the expressions in the literature.
For example, two expressions for ${}^{(2)}\!\bdelta_{\mathrm p}$ with
$\Lambda>0$ have been given. The first, by Tomita~\cite{tom05}, was derived
by solving the perturbation equations at second order in the synchronous-comoving
gauge and then transforming to the Poisson gauge. The second, by Bartolo
\emph{et al}~\cite{baretal10}, was derived by solving the perturbation
equations directly in the Poisson gauge. The two expressions appear to be
completely different. However, by simplifying the ${\mathbb B}$-functions of
Bartolo and introducing our canonical representation of the spatial
dependence we have been able to show, after correcting some typos, that both
of these expressions can be written in our canonical form for
${}^{(2)}\!{\bdelta}_{\mathrm p}$, which is given
by~\eqref{delta_2},~\eqref{newt_special} and~\eqref{A_poisson_spec}.

Third, we have given a systematic procedure for performing a change of gauge
for second order perturbed quantities.  The derivation of our expressions for
${}^{(2)}\!\bdelta_{\bullet}$ relied on solving the perturbation equations in
the Poisson gauge as done in UW, and then using our change of gauge procedure
to calculate ${}^{(2)}\!\bdelta_{\bullet}$ in the other gauges. The procedure
is easy to implement in this application since the change of gauge induces a
simple change in the time-dependent coefficients $A_{i,{\bullet}}$
in~\eqref{delta_2}, while preserving the overall structure of
${}^{(2)}\!\bdelta_{\bullet}$. However, we anticipate that the generality of
our procedure will make it useful in other contexts.

\subsection*{Acknowledgements}
We thank Marco Bruni and David Wands for helpful correspondence concerning
cosmological perturbations and their recent paper~\cite{bruetal14}.  CU also thanks the Department of
Applied Mathematics at the University of Waterloo for kind hospitality. JW
acknowledges financial support from the University of Waterloo.

\begin{appendix}

\section{Spatial differential operators}\label{sec:opdef}

The definitions of the spatial differential operators that we use are as follows. First,
the second order spatial differential operators are defined by
\begin{equation} \label{2order_D}
{\bf D}^2 := \gamma^{ij}{\bf D}_i{\bf D}_j, \qquad {\bf
D}_{ij} := {\bf D}_{(i}{\bf D}_{j)} - \sfrac13 \gamma_{ij}{\bf D}^2,
\end{equation}
where ${\bf D}_i$ denotes covariant differentiation with respect to the
conformal background spatial metric $\gamma_{ij}$. Second, we use the
shorthand notation
\be \label{DA}  \left({\bf D}A\right)^2:=({\bf D}^k A)( {\bf D}_kA), \qquad
{\cal D}(A):={\mathcal S}^{ij} ({\bf D}_ i A)({\bf D}_j A), \ee
where $A$ is a scalar field and ${\mathcal S}^{ij}$ is defined
in~\eqref{modeextractop}. Finally, we define the \emph{mode extraction
operators} (see~\cite{uggwai13b}, equations (85)):
\begin{subequations}\label{modeextractop}
\begin{xalignat}{2}
{\cal S}^{i} &= {\bf D}^{-2}{\bf D}^i ,&\quad {\cal S}^{ij} &= \sfrac32 {\bf
D}^{-2}\!\left({\bf D}^2 + 3K\right)^{-1}{\bf
D}^{ij},\\
{\cal V}_i\!^{j} &= \delta_i\!^j - {\bf D}_i{\cal S}^j, &\quad {\cal
V}_i\!^{jk} &=
\left({\bf D}^2 + 2K\right)^{-1} {\cal V}_i\!^{\la j}{\bf D}^{k\ra},\\
{\cal T}_{ij}\!^{km} &= \delta_i\!^{\la k}\delta_j\!^{m\ra} - {\bf
D}_{(i}{\cal V}_{j)}\!^{km} - {\bf D}_{ij}{\cal S}^{km}. &&
\end{xalignat}
\end{subequations}
If some expression $L({\bf D}_i)$ involving ${\bf D}_i$ scales as $
L(\lambda{\bf D}_i)= {\lambda}^p L({\bf D}_i)$ under a rescaling of
coordinates $x^i\rightarrow {\lambda}^{-1}x^i,\,
\eta\rightarrow{\lambda}^{-1}\eta,$ we say that \emph {$L({\bf D}_i)$ has
weight $p$ in ${\bf D}_i$}. It follows that the canonical differential
expressions in~\eqref{quad_zeta} have the following weights:\footnote {Note
that ${\cal H}\rightarrow {\lambda}{\cal H}$ and $K\rightarrow {\lambda}^2
K.$} $ \zeta^2, \, {\cal D}(\zeta)$ are of weight zero, $ ({\bf D}\zeta)^2,
\, {\bf D}^2{\cal D}(\zeta), \, {\bf D}^2 \zeta^2$ are of weight two and $
{\bf D}^2({\bf D}\zeta)^2, \, {\bf D}^4{\cal D}(\zeta)$ are of weight four.

We now give identities involving the spatial differential operators that we
use to relate results in the literature to our results:
\begin{subequations}  \label{D_iden}
\begin{align}
A{\bf D}^2 A &=\sfrac12{\bf D}^2 A^2 - ({\bf D}A)^2, \label{D1} \\
{\bf D}^i{\bf D}^j \,( {\bf D}_i A{\bf D}_j A) &= \sfrac13{\bf D}^2\left[ ({\bf D}A)^2 + 2({\bf D}^2+3K){\mathcal D}(A) \right],  \label{D5} \\
{\bf D}^i ({\bf D}_iA{\bf D}^2 A) &= -\sfrac16{\bf D}^2\left[({\bf D}A)^2-4({\bf D}^2+3K){\mathcal D}(A)  \right], \label{D4}  \\
({\bf D}^i{\bf D}^j A)\, ({\bf D}_i{\bf D}_j A )&=
\sfrac23\left[({\bf D}^2-3K)({\bf D}A)^2 -({\bf D}^2+3K){\bf D}^2{\mathcal D}(A)\right] +({\bf D}^2A)^2,  \label{D2} \\
{\bf D}^i{\bf D}^j( A\, {\bf D}_i{\bf D}_j A) &=
-\sfrac16{\bf D}^2\left[ 2({\bf D}A)^2+4({\bf D}^2+3K){\mathcal D}(A) -3({\bf D}^2+2K) A^2\right]. \label{D6}
\end{align}
\end{subequations}
%

\section{Transformation laws for gauge invariants} \label{sec:transf}

The first purpose of this appendix is to define the gauge invariants that are
associated with the perturbed metric and matter distribution. We do not,
however, write out the expressions for the gauge invariants in terms of
gauge-variant quantities since our strategy is to work solely with gauge
invariants. First we require the governing equations that determine the gauge
invariants in the Poisson gauge, and these are given in UW. Second we require
a framework for determining how gauge invariants transform under a change of
gauge at second order. For example, given ${}^{(2)}\!{\bdelta}_{\mathrm p}$
(Poisson gauge) how can one calculate ${}^{(2)}\!{\bdelta}_{\mathrm c}$
(uniform curvature gauge) efficiently? The framework that we present in this
appendix is based on the transformation law for the perturbations of a given
tensor field up to second order under a gauge transformation, first given by
Bruni \emph{et al}~\cite{bruetal97}. This transformation law has been used
for this purpose in a number of specific cases (for example,
synchronous-comoving to Poisson~\cite{matetal98, tom05}, Poisson to uniform
curvature~\cite{chretal11}, synchronous-comoving to total
matter~\cite{bruetal14} and Poisson to total matter~\cite{hidetal13}.) Our
goal is to give a general framework that is valid for a specific gauge
invariant and two chosen gauges. In the body of the paper we  consider
pressure-free matter, but in this appendix we assume that the matter content
is a perfect fluid with equation of state $p=w\rho, w=constant$, in order to
increase the applicability of the results.

\subsection{Gauge invariants associated with an arbitrary tensor field} \label{sec:gi}

In cosmological perturbation theory a second order gauge transformation can
be represented in coordinates as follows:
\be {\tilde x}^a = x^a +\epsilon {}^{(1)}\!\xi^a+ \sfrac12
\epsilon^2\left({}^{(2)}\!\xi^a+
{}^{(1)}\!\xi^a\,_{,b}{}^{(1)}\!\xi^b\right), \label{gauge_tf} \ee
where ${}^{(1)}\!\xi^a$ and ${}^{(2)}\!\xi^a$ are independent dimensionless
background vector fields. We consider a 1-parameter family of tensor fields
$A(\epsilon)$, which we assume can be expanded in powers of $\epsilon$, {\it
i.e.} as a Taylor series:
\be\label{taylor} \mathrm{A}(\epsilon) = {}^{(0)}\!\mathrm{A} +
\epsilon\,{}^{(1)}\!\mathrm{A} + \sfrac{1}{2}\epsilon^2\,
{}^{(2)}\!\mathrm{A} + \dots\, ,\ee
where  ${}^{(0)}\!A$ is called the \emph{unperturbed value},
${}^{(1)}\!\mathrm{A}$ is called  the \emph{first order (linear)
perturbation} and ${}^{(2)}\!\mathrm{A}$ is called  the \emph{second order
perturbation} of $A(\epsilon)$. Such a transformation induces a change in the
first and second order perturbations of $A(\epsilon)$ according to
\begin{subequations}   \label{delta_A}
\begin{align}
{}^{(1)}\!{A}[\xi] &= {}^{(1)}\!A +
\pounds_{{}^{(1)}\!\xi}{}^{(0)}\!A,  \label{delta_A1}\\
{}^{(2)}\!{A}[\xi] &= {}^{(2)}\!A +
\pounds_{{}^{(2)}\!\xi}{}^{(0)}\!A +
\pounds_{{}^{(1)}\!\xi}\left(2{}^{(1)}\!A
+ \pounds_{{}^{(1)}\!\xi}{}^{(0)}\!A\right) ,
\label{delta_A2}
\end{align}
\end{subequations}
where $\pounds$ is the Lie derivative (see~\cite{bruetal97}, equations
(1.1)--(1.3)). One fixes a gauge by requiring some components of the
perturbations of some tensor fields ${}^{(r)}\!A[\xi]$, ${}^{(r)}\!B[\xi]$,
etc, with $r=1,2$, to be zero, thereby determining unique values for
${}^{(1)}\!\xi^a$ and ${}^{(2)}\!\xi^a$ which we denote by
${}^{(1)}\!\xi^a_{\bullet}$ and ${}^{(2)}\!\xi^a_{\bullet}$. Since there is
no remaining gauge freedom, the non-zero components ${}^{(r)}\!{\tilde
A}[\xi_{\bullet}]$, obtained by replacing $\xi$ by $\xi_{\bullet}$
in~\eqref{delta_A}, are gauge invariants.  We refer to Malik and
Wands~\cite{malwan09} (see pages 18-20) for an illustration of this process
using the Poisson gauge. When uniquely determined,  the vector fields
${}^{(1)}\!\xi_{\bullet}^a$ and ${}^{(2)}\!\xi_{\bullet}^a$ will be referred
to as \emph{gauge fields}.\footnote{In previous
papers~\cite{uggwai11,uggwai12,uggwai13b} and UW, influenced by the approach
of Nakamura~\cite{nak06,nak07} to cosmological perturbations, we used
the kernel $-X$ to denote a gauge field. Here we use the kernel $\xi$ but
with a subscript, to indicate that the arbitrary vector field $\xi^a$ has
been uniquely determined, thereby fixing a gauge.}

In order to derive a transformation law for gauge invariants under a change
of gauge we consider two gauge fields ${}^{(r)}\!\xi^a_\bullet$ and
${}^{(r)}\!\xi^a_{\mathrm o}$ and define
\begin{subequations}\label{Z}
\begin{align}
{}^{(1)}\!{\bf Z}^a[\xi_\bullet,\xi_{\mathrm o}] &:= {}^{(1)}\!\xi^a_\bullet -
{}^{(1)}\!\xi^a_{\mathrm o},        \label{Z1}\\
{}^{(2)}\!{\bf Z}^a[\xi_\bullet,\xi_{\mathrm o}] &:= {}^{(2)}\!\xi^a_\bullet -
{}^{(2)}\!\xi^a_{\mathrm o} + [{}^{(1)}\!{\xi}_\bullet, {}^{(1)}\!\xi_{\mathrm o}]^a.       \label{Z2}
\end{align}
\end{subequations}
We now consider~\eqref{delta_A} with $\xi=\xi_\bullet$ and $\xi=\xi_{\mathrm
o}$ and form the difference of the two sets of equations. This leads to the
following transformation law relating the gauge invariants
${}^{(r)}\!{A}[\xi_\bullet]$ and ${}^{(r)}\!{A}[\xi_{\mathrm o}]$:
\begin{subequations}\label{Arel}
\begin{align}
{}^{(1)}\!{A}[\xi_\bullet] &= {}^{(1)}\!{A}[\xi_{\mathrm o}] + \pounds_{{}^{(1)}\!{\bf Z}}\,{}^{(0)}\!{A}, \label{Arel1}\\
{}^{(2)}\!{A}[\xi_\bullet] &=
{}^{(2)}\!{A}[\xi_{\mathrm o}] + \pounds_{{}^{(2)}\!{\bf Z}}{}^{(0)}\!{A} + \pounds_{{}^{(1)}\!{\bf Z}}\left( 2{}^{(1)}\!{A}[\xi_{\mathrm o}]
+ \pounds_{{}^{(1)}\!{\bf Z}}{}^{(0)}\!{A} \right).   \label{Arel2}
\end{align}
\end{subequations}
where ${}^{(r)}\!{\bf Z}\equiv {}^{(r)}\!{\bf Z}^a[\xi_\bullet,\xi_{\mathrm
o}]$. We shall refer to the functions ${}^{(r)}\!{\bf
Z}^a[\xi_\bullet,\xi_{\mathrm o}]$, which are gauge invariants, as the
\emph{transition functions.} They are determined by the conditions that
specify the gauge fields ${}^{(r)}\!\xi_\bullet^a$ and
${}^{(r)}\!\xi_{\mathrm o}^a$. We note the formal similarity
between~\eqref{delta_A} and~\eqref{Arel}. In going from~\eqref{delta_A}
to~\eqref{Arel} one replaces gauge-variant quantities by gauge-invariant
quantities: ${}^{(r)}\!{A}[\xi] $ by ${}^{(r)}\!{A}[\xi_\bullet] =
{}^{(r)}\!{A}_\bullet$, ${}^{(r)}\!{A} $ by ${}^{(r)}\!{A}[\xi_{\mathrm o}] =
{}^{(r)}\!{A}_{\mathrm o}$ and ${}^{(r)}\!\xi$ by ${}^{(r)}\!{\bf Z}$.

\subsubsection*{Shorthand notation for gauge invariants and transition functions}

The full notation for the first and second order gauge invariants
associated with a tensor $A(\epsilon)$ is ${}^{(r)}\!{A}[\xi_\bullet],\, r=1,2,$
where $\xi_{\bullet}$ is a gauge field.
If there is no danger of confusion we will use a subscript notation:
\begin{subequations} \label{shorthand}
\be   {}^{(r)}\!{A}_{\bullet} \equiv{}^{(r)}\!{A}[\xi_\bullet]. \ee
The full notation for the transition functions is ${}^{(r)}\!{\bf
Z}^a[\xi_\bullet,\xi_{\mathrm o}]$ where $\xi_\bullet$ and $\xi_{\mathrm o}$
are the two gauge fields. In general we will use the kernel ${\bf Z}$ as
shorthand for ${\bf Z}[\xi_\bullet,\xi_{\mathrm o}]$. If specific gauge
fields are used, for example, $\xi_\bullet=\xi_{\mathrm c}$ and $\xi_{\mathrm
o}=\xi_{\mathrm p}$, we will use subscripts:
\be   {\bf Z}\equiv {\bf Z}[\xi_\bullet,\xi_{\mathrm o}], \qquad
{\bf Z}_{{\mathrm c},{\mathrm p}} \equiv{\bf Z}[\xi_{\mathrm c},\xi_{\mathrm p}].  \ee
\end{subequations}
The source terms in the transformation laws, which have the general form
${\cal F}[{}^{(1)}\!{\bf Z}]$ or ${\mathbb S}[{}^{(1)}\!{\bf Z}]$, are
quadratic in first order gauge invariants.  Specific source terms of the form
${\cal F}[{}^{(1)}\!{\bf Z}_{ {\bullet},{\mathrm p}} ]$ or ${\mathbb
S}[{}^{(1)}\!{\bf Z}_{{\bullet},{\mathrm p}}]$ are quadratic in the first
order gauge invariants ${}^{(1)}\!\Psi_{\mathrm p},
{}^{(1)}\!{\bdelta}_{\mathrm p}$ and ${}^{(1)}\!{\bf v}_{\mathrm p}$. We will
omit the superscript $ {}^{(1)}$ and the subscript $_{\mathrm p}$ when there
is no danger of confusion.

\subsection{The metric gauge invariants} \label{sec:metric}

The gauge invariants ${}^{(r)}\!g_{ab}[\xi]$ associated with the metric
$g_{ab}$ are given by~\eqref{delta_A} with the arbitrary tensor $A$ chosen to
be $g_{ab}$. Since $a^{-2}g_{ab}$ is dimensionless we can define
dimensionless gauge invariants by
\be {}^{(r)}\!{\bf f}_{ab}[\xi]:=a^{-2}\,{}^{(r)}\!g_{ab}[\xi]. \label{bold_f} \ee
We choose the tensor $A$ in equation~\eqref{Arel} to be $g_{ab}$ and
use~\eqref{bold_f} to obtain the
following transformation law for ${}^{(r)}\!{\bf f}_{ab}[\xi]$:
\begin{subequations}\label{bold_f_transf}
\begin{align}
{}^{(1)}\!{\bf f}_{ab}[\xi_\bullet] &=
{}^{(1)}\!{\bf f}_{ab}[\xi_{\mathrm o}] + a^{-2}\pounds_{{}^{(1)}\!{\bf Z}}\!\left(a^2\gamma_{ab}\right), \label{bold_f1}\\
{}^{(2)}\!{\bf f}_{ab}[\xi_\bullet] &=
{}^{(2)}\!{\bf f}_{ab}[\xi_{\mathrm o}] + a^{-2}\pounds_{{}^{(2)}\!{\bf Z}}\!\left(a^2\gamma_{ab}\right) +
{\mathcal F}_{ab}[{}^{(1)}{\bf Z}],    \label{bold_f2}
\end{align}
where
\be {\mathcal F}_{ab}[{}^{(1)}\!{\bf Z}] := a^{-2}\pounds_{{}^{(1)}\!{\bf
Z}}\!\left(2a^2 \,{}^{(1)}\!{\bf f}_{ab}[\xi_{\mathrm o}] +
\pounds_{{}^{(1)}\!{\bf Z}}\!\left(a^2\gamma_{ab}\right)\right)\!.
\label{F2X2} \ee
\end{subequations}
Here $\gamma_{ab}$ is the conformally related background metric, given by
${}^{(0)}\!g_{ab}=a^2\gamma_{ab}$.

We now perform a mode decomposition of ${}^{(r)}\!{\bf f}_{ab}[\xi]$ as
follows:\footnote{We use notation that is compatible with the notation
in~\cite{uggwai13b}. See equations (24) and (88).}
\begin{subequations} \label{bold_f_split}
\begin{align}
{}^{(r)}\!{\bf f}_{00}[\xi] &= -2  {}^{(r)}\!\Phi[\xi] \, ,\\
{}^{(r)}\!{\bf f}_{0 i}[\xi] &= {\bf D}_i {}^{(r)}\!{\bf B}[\xi] + {}^{(r)}\!{\bf B}_i[\xi]\, ,\\
{}^{(r)}\!{\bf f}_{ij}[\xi] &= -2{}^{(r)}\!\Psi[\xi] \gamma_{ij}  + 2{\bf D}_i {\bf D}_j {}^{(r)}\!{\bf C}[\xi] +
2{\bf D}_{(i} {}^{(r)}\!{\bf C}_{j)}[\xi]+ 2{}^{(r)}\!{\bf C}_{ij}[\xi]\,.
\end{align}
\end{subequations}
We can apply the mode extraction operators defined in
equations~\eqref{modeextractop} in appendix~\ref{sec:opdef}
to~\eqref{bold_f_transf} to obtain the transformation laws for the individual
gauge invariants on the right side of~\eqref{bold_f_split}, obtaining\footnote
{Here ${\eta}$ is conformal time, and $\partial_{\eta}={\cal H}x\partial_x.$}
\begin{subequations}\label{modefXrel}
\begin{align}
{}^{(r)}\!\Phi[\xi_\bullet] &= {}^{(r)}\!\Phi[\xi_{\mathrm o}] +
(\partial_\eta + {\cal H}){}^{(r)}\!{\bf Z}^0 -\sfrac12
{\mathcal F}_{00}[{\bf Z}], \label{relPhiX}\\
{}^{(r)}\!{\bf B}[\xi_\bullet] &= {}^{(r)}\!{\bf B}[\xi_{\mathrm o}] - {}^{(r)}\!{\bf Z}^0 + \partial_\eta{}^{(r)}\!{\bf Z}
+ {\cal S}^{i}{\mathcal F}_{0i}[{\bf Z}], \label{relBX}\\
{}^{(r)}\!\Psi[\xi_\bullet] &= {}^{(r)}\!\Psi[\xi_{\mathrm o}] - {\cal H}{}^{(r)}\!{\bf Z}^0 -
\sfrac16({\mathcal F}^k\!_k - {\bf D}^2{\cal S}^{ij}\hat{{\mathcal F}}_{ij})[{\bf Z}], \label{relPsiX}  \\
{}^{(r)}\!{\bf B}_i[\xi_\bullet] &= {}^{(r)}\!{\bf B}_i[\xi_{\mathrm o}] + \partial_\eta{}^{(r)}\!\tilde{\bf Z}_i
+ {\cal V}_i\!^{j}{\mathcal F}_{0j}[{\bf Z}], \label{relBiX}\\
{}^{(r)}\!{\bf C}[\xi_\bullet] &= {}^{(r)}\!{\bf C}[\xi_{\mathrm o}] + {}^{(r)}\!{\bf Z}
+ \sfrac12\,{\cal S}^{ij} \hat{{\mathcal F}}_{ij}[{\bf Z}], \label{relCX}\\
{}^{(r)}\!{\bf C}_i[\xi_\bullet] &= {}^{(r)}\!{\bf C}_i[\xi_{\mathrm o}] + {}^{(r)}\!\tilde{\bf Z}_i +
{\cal V}_i\!^{jk}\hat{{\mathcal F}}_{jk}[{\bf Z}], \label{relCiX}\\
{}^{(r)}\!{\bf C}_{ij}[\xi_\bullet] &= {}^{(r)}\!{\bf C}_{ij}[\xi_{\mathrm o}] +
\sfrac12\,{\cal T}_{ij}\!^{km}\hat{{\mathcal F}}_{km}[{\bf Z}], \label{relCijX}
\end{align}
\end{subequations}
where $r=1,2$. The source terms ${\mathcal F}_{ab}[{\bf Z}]$, which are given
by~\eqref{F2X2},  do not appear when $r=1$. We will give explicit expressions
for them later. Here we have decomposed the transition functions
${}^{(r)}\!{\bf Z}^a\equiv{}^{(r)}\!{\bf Z}^a[\xi_\bullet, \xi_{\mathrm o}]$
according to
\be {}^{(r)}\!{\bf Z}^a = ({}^{(r)}\!{\bf Z}^0,{}^{(r)}\!{\bf Z}^i), \qquad
{}^{(r)}\!{\bf Z}^i={\bf D}^i{}^{(r)}\!{\bf Z} + {}^{(r)}\!\tilde{{\bf Z}}^i,
\qquad {\bf D}_i{}^{(r)}\!\tilde{{\bf Z}}^i=0.  \ee
%

%
%

\subsection{Density gauge invariants}  \label{sec:density}

We choose  $A=\rho$, the matter density scalar~ in eq.~\eqref{Arel}. On
evaluating the Lie derivatives we obtain
\begin{subequations}\label{scalartransf}
\begin{align}
{}^{(1)}\!\rho[\xi_\bullet] &= {}^{(1)}\!\rho[\xi_{\mathrm o}] + {}^{(1)}\!{\bf Z}^0\,{}^{(0)}\!\rho^\prime ,\\
{}^{(2)}\!\rho[\xi_\bullet] &= {}^{(2)}\!\rho[\xi_{\mathrm o}] + {}^{(2)}\!{\bf Z}^0\,{}^{(0)}\!\rho^\prime +
\left({}^{(1)}\!{\bf Z}^0\partial_\eta + {}^{(1)}\!{\bf Z}^i{\bf D}_i\right)\left(2 {}^{(1)}\!\rho[\xi_{\mathrm o}]
+ {}^{(1)}\!{\bf Z}^0\,{}^{(0)}\!\rho^\prime\right),
\end{align}
\end{subequations}
where $'$ denotes differentiation with respect to conformal time ${\eta}$.
Here and in the rest of this section the kernel ${\bf Z}$ is shorthand for
${\bf Z}[\xi_\bullet,\xi_{\mathrm o}]$. We introduce dimensionless gauge
invariants by normalizing with the inertial mass density:
\begin{equation}
{}^{(r)}\!{\bdelta}[\xi] := \frac{{}^{(r)}\!\rho[\xi]}{{}^{(0)}\!\rho
+ {}^{(0)}\!p},
\end{equation}
which in the case of dust is just the usual fractional density perturbation.
Then~\eqref{scalartransf} leads to the following transformation law for the
density gauge invariants:
\begin{subequations}\label{delta_transf}
\begin{align}
{}^{(1)}\!{\bdelta}_{\bullet} &= {}^{(1)}\!{\bdelta}_{\mathrm o} - 3{\cal H}{}^{(1)}\!{\bf Z}^0,\\
{}^{(2)}\!{\bdelta}_{\bullet}&= {}^{(2)}\!{\bdelta}_{\mathrm o} - 3{\cal H}{}^{(2)}\!{\bf Z}^0 + {\cal F}_\delta[{}^{(1)}\!{\bf Z}],
\end{align}
where
\begin{equation}  \label{delta_source}
{\cal F}_\delta[{}^{(1)}\!{\bf Z}] := \left({\bf Z}^0(\partial_\eta - 3(1 +
w){\cal H}) + {\bf Z}^i {\bf D}_i\right)\!\left(2{\bdelta}[\xi_{\mathrm o}] - 3{\cal
H}{\bf Z}^0\right),
\end{equation}
\end{subequations}
and we are using the shorthand notation~\eqref{shorthand}. Here we have
dropped the superscript ${}^{(1)}$ on the first order quantities on the right
hand side of this equation. In deriving equations~\eqref{delta_transf} we
used the following background equations for a perfect fluid:
\be {}^{(0)}\!\rho^\prime=-3{\cal H}( {}^{(0)}\!\rho+ {}^{(0)}\!p), \quad
{}^{(0)}\! p^\prime=w\,{}^{(0)}\!\rho^\prime,  \ee
where $w$ is the constant equation of state parameter.

\subsection{Velocity gauge invariants}  \label{sec:velocity}

The gauge invariants ${}^{(r)}\!u_{a}[\xi]$ associated with the covariant
unit vector field $u_{a}$ are given by~\eqref{delta_A} with the arbitrary
tensor $A$ chosen to be $u_{a}$. Since $a^{-1}u_{a}$ is dimensionless we can
define dimensionless gauge invariants by
\be {}^{(r)}\!{\bf v}_{a}[\xi]:=a^{-1}{}^{(r)}\!u_{a}[\xi]. \label{bold_v} \ee
Equation~\eqref{Arel}, with the tensor $A$ chosen to be $u_{a}$, in
conjunction with~\eqref{bold_v}, then leads to the following transformation
law for ${}^{(r)}\!{\bf v}_{a}[\xi_{\mathrm o}]$:
\begin{subequations}\label{bold_v_transf}
\begin{align}
{}^{(1)}\!{\bf v}_{a}[\xi_\bullet] &:=
{}^{(1)}\!{\bf v}_{a}[\xi_{\mathrm o}] + a^{-1}\pounds_{{}^{(1)}\!{\bf Z}}\!\left(a {}^{(0)}v_{a}\right), \label{bold_v1}\\
{}^{(2)}\!{\bf v}_{a}[\xi_\bullet] &:=
{}^{(2)}\!{\bf v}_{a}[\xi_{\mathrm o}] + a^{-1}\pounds_{{}^{(2)}\!{\bf Z}}\!\left(a{}^{(0)}v_{a}\right) +
({\mathcal F}_{v})_a[{}^{(1)}\!{\bf Z}],    \label{bold_v2}
\end{align}
where
\be ({\mathcal F}_{v})_a[{}^{(1)}\!{\bf Z}] := a^{-1}\pounds_{{}^{(1)}\!{\bf
Z}}\!\left(2a \,{}^{(1)}\!{\bf v}_{a}[\xi_{\mathrm o}] +
\pounds_{{}^{(1)}\!{\bf Z}}\!\left(a {}^{(0)}\!v_{a}\right)\right)\!,
\label{new} \ee
\end{subequations}
and $a {}^{(0)}v_{a}\equiv {}^{(0)}u_{a}$. Evaluating the Lie derivatives and
restricting to the spatial components yields the following
\begin{subequations}\label{vectortransf}
\begin{align}
{}^{(1)}\!{\bf v}_i[\xi_\bullet] &= {}^{(1)}\!{\bf v}_i[\xi_{\mathrm o}]  - {\bf D}_i{}^{(1)}\!{\bf Z}^0,\\
{}^{(2)}\!{\bf v}_i[\xi_\bullet] &=
{}^{(2)}\!{\bf v}_i[\xi_{\mathrm o}]  - {\bf D}_i{}^{(2)}\!{\bf Z}^0 + ({\cal F}_\mathrm{v})_i[{}^{(1)}\!{\bf Z}],
\end{align}
where
\be
\begin{split}
({\cal F}_\mathrm{v})_i[{}^{(1)}\!{\bf Z}] &\!:= 2{\bf Z}^0(\partial_\eta + {\cal H}){\bf v}_i[\xi_{\mathrm o}]  -
2\Phi[\xi_{\mathrm o}]{\bf D}_i{\bf Z}^0 - \sfrac12{\bf D}_i(\partial_\eta+2{\cal
H})({\bf Z}^0)^2\\
& \quad - {\bf D}_i({\bf Z}^j{\bf D}_j {\bf Z}^0) +
2\left({\bf Z}^j{\bf D}_j{\bf v}_i[\xi_{\mathrm o}]   + ({\bf D}_i {\bf Z}^j){\bf v}_j[\xi_{\mathrm o}] \right).
\end{split}
\ee
\end{subequations}
We now mode decompose ${}^{(r)}\!{\bf v}_i[\xi]$ into a scalar and vector
part according to ${\bf v}_i = {\bf D}_i {\bf v} + \tilde{{\bf v}}_i$, ${\bf
D}^i\tilde{{\bf v}}_i=0$. On restricting to the purely scalar case at linear
order (\emph{i.e.} ${\bf v}_i={\bf D}_i{\bf v},\, {\bf Z}^i={\bf D}^i{\bf
Z}$),~\eqref{vectortransf} reduces to\footnote{Apply the
mode extraction operator ${\mathcal S}^i$ to the
second of equations~\eqref{vectortransf} to get the second of
equations~\eqref{svectortransfrel}. We introduce the shorthand notation
${\cal F}_\mathrm{v}\equiv {\cal S}^i({\cal F}_\mathrm{v})_i.$ }
\begin{subequations}\label{svectortransfrel}
\begin{align}
{}^{(1)}\!{\bf v}_{\bullet} &= {}^{(1)}\!{\bf v}_{\mathrm o}  - {}^{(1)}\!{\bf Z}^0,\\
{}^{(2)}\!{\bf v}_{\bullet} &= {}^{(2)}\!{\bf v}_{\mathrm o}  - {}^{(2)}\!{\bf Z}^0 + {\cal F}_\mathrm{v}[{}^{(1)}\!{\bf Z}],
\end{align}
where
\begin{equation}  \label{vec_source}
\begin{split}
{\cal F}_\mathrm{v}[{}^{(1)}\!{\bf Z}] &= 2{\cal
S}^i\left({\bf Z}^0 (\partial_\eta + {\cal H}){\bf D}_i{\bf v}[\xi_{\mathrm o}] -
\Phi[\xi_{\mathrm o}]{\bf D}_i{\bf Z}^0\right)\\
& \quad - \sfrac12(\partial_\eta + 2{\cal
H})({\bf Z}^0)^2 - ({\bf D}^j {\bf Z}){\bf D}_j({\bf Z}^0 - 2{\bf v}[\xi_{\mathrm o}]),
\end{split}
\end{equation}
\end{subequations}
and we are using the shorthand notation~\eqref{shorthand}.

\subsection{Transformation laws between the Poisson, the uniform curvature and the total
matter gauges}\label{sec:poisson}

The Poisson, uniform curvature and total matter gauges all satisfy the following
conditions on the metric gauge invariants:
\be {}^{(r)}\!{\bf C}[\xi]=0, \qquad {}^{(r)}\!{\bf C}_i[\xi]=0,
\label{spatX}  \ee
for $r=1,2$, where $\xi$ is any of the gauge fields $\xi_{\mathrm
p},\xi_{\mathrm c}$ and $\xi_{\mathrm v}$. It follows from~\eqref{relCX}
and~\eqref{relCiX} with $r=1$ that the spatial part of the first order
transition function ${}^{(1)}\!{\bf Z}^a$ relating these three gauges will be
zero:
\be  {}^{(1)}\!{\bf Z}[\xi_\bullet, \xi] = 0, \qquad  {}^{(1)}\!\tilde{\bf
Z}_i[\xi_\bullet, \xi] =0, \label{spatZ} \ee
where $\xi_\bullet$ and $\xi$ can be chosen to be any two of the gauge fields
$\xi_{\mathrm p},\xi_{\mathrm c}$ and $\xi_{\mathrm v}$. On the other hand,
these three gauges are distinguished by the specification of the temporal
gauge, as follows:
\be {}^{(r)}\!{\bf B}[\xi_{\mathrm p}]=0, \qquad {}^{(r)}\!\Psi[\xi_{\mathrm
c}]=0, \qquad   {}^{(r)}\!{\bf v}[\xi_{\mathrm v}]=0, \label{temp_gauge}
\ee
for $r=1,2$ respectively.

We now give the components of the source terms ${\mathcal F}_{ab}[{\bf Z}]$
in the transformation laws~\eqref{modefXrel}, assuming that the linear metric
perturbation is purely scalar, and that the conditions~\eqref{spatX} and hence~\eqref{spatZ} are
satisfied. We calculate the Lie derivatives in~\eqref{F2X2}, making use
of~\eqref{bold_f_split},~\eqref{spatX} and~\eqref{spatZ}, which leads to
\begin{subequations}\label{metric_source}
\begin{align}
{\mathcal F}_{00}[{}^{(1)}\!{\bf Z}] &= -2\!\left[{\bf Z}^0(\partial_{\eta} + 2{\cal H}) +
2(\partial_\eta {\bf Z}^0)\right]\!\left[(\partial_\eta + {\cal H}){\bf Z}^0 + 2\Phi[\xi]\right]\!,\\
\begin{split}
{\mathcal F}_{0i}[{}^{(1)}\!{\bf Z}] &= -\left[{\bf Z}^0(\partial_{\eta} + 2{\cal H}) +
(\partial_\eta {\bf Z}^0)\right]\!{\bf D}_i({\bf Z}^0 - 2{\bf B}[\xi]) \\
& \quad - 2({\bf D}_i {\bf Z}^0)\left((\partial_\eta + {\cal H}){\bf Z}^0 + 2\Phi[\xi]\right)\!, \end{split}\\
{\mathcal F}^k\!_k[{}^{(1)}\!{\bf Z}] &=
6{\bf Z}^0(\partial_{\eta} + 2{\cal H})({\cal H}{\bf Z}^0 - 2\Psi[\xi]) - 2({\bf D}^k {\bf Z}^0){\bf D}_k({\bf Z}^0 - 2{\bf B}[\xi]),\\
\hat{\mathcal F}_{ij}[{}^{(1)}\!{\bf Z}] &= -2{\bf D}_{\la i}({\bf Z}^0 - 2{\bf B}[\xi])({\bf D}_{j\ra} {\bf Z}^0),
\end{align}
\end{subequations}
where ${\bf Z}^0\equiv{\bf Z}^0[\xi_\bullet,\xi]$, and $\xi_\bullet$ and
$\xi$ can be chosen to be any two of the gauge fields $\xi_{\mathrm
p},\xi_{\mathrm c}$ and $\xi_{\mathrm v}$.

When evaluating the source terms~\eqref{delta_source},~\eqref{vec_source}
and~\eqref{metric_source} in the following sections it is convenient to
eliminate the temporal derivatives of the first order gauge invariants
${\bdelta}, {\bf v}$ and ${\Psi}$ in the Poisson gauge. To do this we use the
linearized conservation equations for a perfect fluid in the following
form:\footnote{Choose ${\xi}=\xi_{\mathrm p}$ in equations (43) in~\cite{uggwai12}, and
specialize to a perfect fluid by setting ${\bar \Gamma}=0, {\bar \Xi}=0$ and
noting that $\Phi_{\mathrm p}=\Psi_{\mathrm p}$. Also note that ${\mathbb
V}_{\mathrm p}={\bf v}_{\mathrm p}$ and ${\mathbb D}\equiv {\bdelta}_{\mathrm
v}={\bdelta}_{\mathrm p} -3{\cal H}{\bf v}_{\mathrm p}.$}
%
\begin{subequations} \label{evol}
\begin{align} x\,\partial_x({\bdelta}_{\mathrm p}-3\Psi_{\mathrm p}) +{\cal H}^{-2}{\bf D}^2({\cal H}{\bf v}_{\mathrm p}) &= 0, \label{evol1} \\
{\cal H}(x\,\partial_x +1){\bf v}_{\mathrm p} +\Psi_{\mathrm p} +
w\,({\bdelta}_{\mathrm p} -3{\cal H}{\bf v}_{\mathrm p}) &=0, \label{evol2}
\end{align}
and  the velocity equation in the form
\be (x\,\partial_x +1)\Psi_{\mathrm p} = -\frac{\cal A}{{2\cal H}^2}({\cal H}{\bf v}_{\mathrm p}),
\label{evol3}  \ee
\end{subequations}
(see equations (15b) and (16b) in UW). Here the scalar ${\cal A}$ is given by
\be {\cal A}=2(-\partial_{\eta}{\cal H}+{\cal H}^2+K) = 3(1+w){\cal H}^2 \Omega_m, \label{calA} \ee
the second equality holding for a perfect fluid with linear equation of
state.\footnote{See equations (16) and (36) in~\cite{uggwai12}, noting that
the background Einstein equations imply ${\cal A}_G={\cal A}_T$, or equation
(6) in UW, but note the typo: the signs on ${\cal H}'$ and ${\cal H}^2$ are
reversed.} In addition~\eqref{evol2} and~\eqref{calA} lead to
\be x\,\partial_x({\cal H}{\bf v}_{\mathrm p})= -\frac{{\cal A}-2K}{2{\cal
H}^2}({\cal H}{\bf v}_{\mathrm p}) -\Psi_{\mathrm p} - w\,
({\bdelta}_{\mathrm p} -3{\cal H}{\bf v}_{\mathrm p}), \label{evol2a} \ee
on noting that $\partial_{\eta}={\cal H}x\partial_x$.

\subsubsection{Transforming from the Poisson to the uniform curvature gauge}

The transition quantities ${}^{(r)}\!{\bf Z}^0_{{\mathrm c},{\mathrm p}}
\equiv{}^{(r)}\!{\bf Z}^0[\xi_{\mathrm c},\xi_{\mathrm p}] $ are obtained by
choosing $\xi_\bullet=\xi_{\mathrm c}$ and $\xi_{\mathrm o}=\xi_{\mathrm p}$
in~\eqref{relPsiX} and using the second of equations~\eqref{temp_gauge}. This
leads to
\begin{subequations}   \label{svectortransfCP}
\begin{align}
{\cal H}{}^{(1)}\!{\bf Z}^0_{{\mathrm c},{\mathrm p}} &= {}^{(1)}\!{\Psi}_\mathrm{p},  \label{svectortransfCP1}  \\
{\cal H}{}^{(2)}\!{\bf Z}^0_{{\mathrm c},{\mathrm p}} &= {}^{(2)}\!{\Psi}_\mathrm{p}  -
\sfrac16\left({\mathcal F}^k\!_k - {\bf D}^2{\cal S}^{ij}\hat{{\mathcal F}}_{ij}\right)[{}^{(1)}\!{\bf Z}_{{\mathrm c},{\mathrm p}}].
\end{align}
\end{subequations}
Next, we substitute~\eqref{svectortransfCP} in~\eqref{delta_transf} with $_{\bullet}$ and $_{\mathrm o}$
replaced by $_{\mathrm c}$ and $_{\mathrm p}$, and obtain
\begin{subequations}
\begin{align}
{}^{(1)}\!{\bdelta}_{\mathrm c} &= {}^{(1)}\!{\bdelta}_{\mathrm p} - 3{}^{(1)}\!{\Psi}_{\mathrm p} ,\label{ptoc1} \\
{}^{(2)}\!{\bdelta}_{\mathrm c} &= {}^{(2)}\!{\bdelta}_{\mathrm p} - 3{}^{(2)}\!{\Psi}_{\mathrm p} +
 {\mathbb S}_\delta[{}^{(1)}\!{\bf Z}_{{\mathrm c},{\mathrm p}}  ], \label{ptoc2}
\end{align}
where
\be  {\mathbb S}_\delta[{}^{(1)}\!{\bf Z}_{{\mathrm c},{\mathrm p}}]:= \sfrac12(2{\cal F}_\delta
 + {\mathcal F}^k\!_k - {\bf
D}^2{\cal S}^{ij}\hat{{\mathcal F}}_{ij})[{}^{(1)}\!{\bf Z}_{{\mathrm c},{\mathrm p}}] .
\ee
\end{subequations}
We now use~\eqref{delta_source},~\eqref{metric_source}, the first of
equations~\eqref{temp_gauge}, and~\eqref{svectortransfCP1} to show that
\be  {\mathbb S}_\delta[{}^{(1)}\!{\bf Z}_{{\mathrm c},{\mathrm p}}]=
3\Psi[(1 + 3w)\Psi-2(1+w){\bdelta}] + 2\Psi x\,\partial_x({\bdelta}-3\Psi)-
{\cal H}^{-2}\left[({\bf D}\Psi)^2-{\bf D}^2 {\cal D}(\Psi) \right],  \ee
where we for brevity drop the subscript $_\mathrm{p}$ on the first order
quantities in the source terms. Finally we use~\eqref{evol1} to eliminate the
temporal derivative, obtaining
\be  {\mathbb S}_\delta[{}^{(1)}\!{\bf Z}_{{\mathrm c},{\mathrm p}}]=
3\Psi[(1 + 3w)\Psi-2(1+w){\bdelta}] - {\cal H}^{-2}\left[2\Psi{\bf D}^2({\cal
H}{\bf v}) + ({\bf D}\Psi)^2-{\bf D}^2 {\cal D}(\Psi)  \right].
\label{source_cp} \ee
In summary, equation~\eqref{ptoc2}, with the source term given
by~\eqref{source_cp}, is the transformation law that relates
${}^{(2)}\!{\bdelta}_{\mathrm c}$ to ${}^{(2)}\!{\bdelta}_{\mathrm p}$.

\subsubsection{Transforming from the Poisson to the total matter gauge}

The transition quantities ${}^{(r)}\!{\bf Z}^0_{{\mathrm v},\,{\mathrm p}}
\equiv{}^{(r)}\!{\bf Z}^0[\xi_{\mathrm v},\xi_{\mathrm p}] $ are obtained by
replacing $_{\bullet}$ and $_{\mathrm o}$ with $_{\mathrm v}$ and $_{\mathrm
p}$ in~\eqref{svectortransfrel} and using the third of
equations~\eqref{temp_gauge}. This leads to
\begin{subequations}\label{svectortransfPT}
\begin{align}
{}^{(1)}\!{\bf Z}^0_{{\mathrm v},{\mathrm p}} &= {}^{(1)}\!{\bf v}_\mathrm{p}, \label{svectortransfPT1}  \\
{}^{(2)}\!{\bf Z}^0_{{\mathrm v},{\mathrm p}} &= {}^{(2)}\!{\bf v}_\mathrm{p} +
{\cal F}_\mathrm{v}[{}^{(1)}\!{\bf Z}_{{\mathrm v},{\mathrm p}}], \label{svectortransfPT2}
\end{align}
\end{subequations}
It follows from~\eqref{delta_transf} that ${}^{(r)}\!{\bdelta}_{\mathrm v}$
is related to ${}^{(r)}\!{\bdelta}_{\mathrm p}$
according to
\begin{subequations}
\begin{align}
{}^{(1)}\!{\bdelta}_{\mathrm v} &= {}^{(1)}\!{\bdelta}_{\mathrm p} - 3{\cal H}{}^{(1)}\!{\bf v}_{\mathrm p}  , \label{ptov1} \\
{}^{(2)}\!{\bdelta}_{\mathrm v} &= {}^{(2)}\!{\bdelta}_{\mathrm p} - 3{\cal H}{}^{(2)}\!{\bf v}_{\mathrm p} +
{\mathbb S}_\delta[{}^{(1)}\!{\bf Z}_{{\mathrm v},{\mathrm p}}],  \label{ptov2}
\end{align}
where
\be {\mathbb S}_\delta[{}^{(1)}\!{\bf Z}_{{\mathrm v},{\mathrm p}}]:=( {\cal F}_\delta
 -3{\cal H}{\cal F}_\mathrm{v})[{}^{(1)}\!{\bf Z}_{{\mathrm v},{\mathrm p}}].   \ee
\end{subequations}
Then we use~\eqref{delta_source},~\eqref{vec_source}, the first of equations~\eqref{temp_gauge},
and~\eqref{svectortransfPT1} to calculate an explicit expression for the source term:
\begin{subequations}
\begin{align}
{\cal F}_\delta[{}^{(1)}\!{\bf Z}_{{\mathrm v},{\mathrm p}}] &=
{\cal H}{\bf v}\left(x\,\partial_x - 3(1+w)\right) (2{\bdelta} - 3{\cal H}{\bf v}),  \\
{\cal F}_\mathrm{v}[{}^{(1)}\!{\bf Z}_{{\mathrm v},{\mathrm p}}] &= 2{\cal
S}^i[{\cal H}{\bf v}\,{\bf D}_i (x\,\partial_x {\bf v}) - \Phi {\bf D}_i
{\bf v}] - {\cal H}{\bf v}x\,\partial_x{\bf v}.
\end{align}
\end{subequations}
Eliminating the time derivatives using equations~\eqref{evol} and~\eqref{evol2a}
and making use of~\eqref{omega_def} and~\eqref{calA} we
obtain
\be  \begin{split}
{\mathbb S}_\delta[{}^{(1)}\!{\bf Z}_{{\mathrm v},{\mathrm p}}]= &{\cal H}{\bf v}\left[-6{\bdelta} +
3\left(3-\sfrac32(1+w)\Omega_m+ \Omega_k\right){\cal H}{\bf v}- 2{\cal H}^{-2} {\bf D}^2({\cal H}{\bf v})\right] \\
& -6w\,{\cal S}^i({\bdelta}\,{\bf D}_i ({\cal H}{\bf v})).
\end{split}   \label{source_vp}
 \ee
In summary, equation~\eqref{ptov2}, with the source term given
by~\eqref{source_vp}, is the transformation law that relates
${}^{(2)}\!{\bdelta}_{\mathrm v}$ to ${}^{(2)}\!{\bdelta}_{\mathrm p}$.

\subsection{Transforming from the Poisson to the uniform density gauge}

The uniform density gauge is defined by
\be {}^{(r)}\!{\bdelta}[\xi_{\rho}]=0,  \label{udg}   \ee
with the spatial part of the gauge field  fixed as for the Poisson gauge
in~\eqref{spatX}. We specialize~\eqref{relPsiX} by choosing $\xi_\bullet
=\xi_{\rho}$ and $\xi=\xi_{\mathrm p}$, which relates ${}^{(r)}\!\Psi_{\rho}$
to ${}^{(r)}\!\Psi_{\mathrm p}$. Substituting equation~\eqref{udg}
in~\eqref{delta_transf} with $_{\bullet}$ and $_{\mathrm o}$ replaced by
$_{\rho}$ and $_{\mathrm p}$, yields expressions for the required transition
quantities ${}^{(r)}\!{\bf Z}^0_{{\rho},\,{\mathrm p}} \equiv{}^{(r)}\!{\bf
Z}^0[\xi_{\rho},\xi_{\mathrm p}] $, which when substituted in~\eqref{relPsiX}
lead to
\begin{subequations} \label{Psi_rho}
\begin{align}
{}^{(1)}\!\Psi_{\rho} &= {}^{(1)}\!\Psi_{\mathrm p}-\sfrac13 {}^{(1)}\!{\bdelta}_{\mathrm p}, \\
{}^{(2)}\!\Psi_{\rho} &= {}^{(2)}\!\Psi_{\mathrm p}-\sfrac13 {}^{(2)}\!{\bdelta}_{\mathrm p} + {\mathbb S}_{\Psi}[{\bf Z}_{\rho,\mathrm p}],
\end{align}
where
\be  {\mathbb S}_{\Psi}[{\bf Z}_{\rho,\mathrm p}]:= -\sfrac16\left(2{\cal F}_\delta  +
{\mathcal F}^k\!_k - {\bf D}^2{\cal S}^{ij}\hat{{\mathcal F}}_{ij}\right)[{\bf Z}_{\rho,\mathrm p}].\ee
Finally it follows from~\eqref{delta_source},~\eqref{metric_source} and~\eqref{evol1} that
\be  {\mathbb S}_{\Psi}[{\bf Z}_{\rho,\mathrm p}]= \sfrac19
{\bdelta}\left((1+3w){\bdelta} +12\Psi\right) +\sfrac{1}{27}{\cal H}^{-2}\left( ({\bf D}
{\bdelta})^2-{\bf D}^2 {\mathcal D}({\bdelta}) + 6{\bdelta} {\bf D}^2 {\cal
H}{\bf v}\right). \ee
\end{subequations}
%

\subsection{Transforming from the total matter to the synchronous-comoving gauge} \label{sec:synch}

In cosmological perturbation theory when considering perturbations of FL
universes containing pressure-free matter that is \emph{irrotational} one can
specialize to the \emph{synchronous-comoving gauge}.\footnote{See for
example~\cite{jeoetal12},~\cite{bruetal12}, and~\cite{hwanoh06}.  The last
reference discusses the relation between the synchronous-comoving gauge and
the traditional synchronous gauge.}

In our notation this gauge
is defined by the following conditions:
\be {}^{(r)}\!{\bf B}[\xi_\mathrm{s}]=0, \qquad
{}^{(r)}\!\Phi[\xi_\mathrm{s}]=0, \qquad {}^{(r)}\!{\bf
v}[\xi_\mathrm{s}]=0,\qquad r=1,2, \label{sg} \ee
where the subscript $_\mathrm{s}$ stands for synchronous-comoving. We have to
determine the transition quantities ${}^{(r)}\!{\bf Z}^0_{{\mathrm
s},{\mathrm v}} \equiv{}^{(r)}\!{\bf Z}^0[\xi_{\mathrm s},\xi_{\mathrm v}] $
and ${}^{(1)}\!{\bf Z}_{{\mathrm s},{\mathrm v}} \equiv{}^{(1)}\!{\bf
Z}[\xi_{\mathrm s},\xi_{\mathrm v}] $, where the subscript $_{\mathrm{v}}$ refers to
the total matter gauge. First, since $ {}^{(r)}\!{\bf
v}[\xi_\mathrm{v}]=0,$ it follows from~\eqref{sg}
and~\eqref{svectortransfrel} with $\xi_\bullet=\xi_\mathrm{s}$ and
$\xi=\xi_\mathrm{v}$ that the temporal parts are zero:
\be {}^{(1)}\!{\bf Z}^0_{{\mathrm s},{\mathrm v}}=0, \qquad {}^{(2)}\!{\bf
Z}^0_{{\mathrm s},{\mathrm v}} =0,  \label{Xvp1}  \ee
unlike in the previous cases. The spatial part is determined as follows.
Since ${}^{(1)}\!{\bf B}[\xi_\mathrm{s}] = 0$ equation~\eqref{relBX} with
$r=1$ leads to
\begin{equation}
x{\cal H}\,\partial_x({}^{(1)}{\bf Z}_{{\mathrm s},{\mathrm v}}) = -{}^{(1)}\!{\bf B}_\mathrm{v}
 = {}^{(1)}\!{\bf v}_\mathrm{p}.  \label{Xvp2}
\end{equation}
Noting that in the case of dust we have ${\cal A}x=3m^2$, which follows
from~\eqref{omega_m} and~\eqref{calA}, we can write~\eqref{evol3} in the form
\be  {}^{(1)}\!{\bf v}_\mathrm{p}= - x{\cal H}\left(\sfrac23
m^{-2}\partial_x(x ^{(1)}\Psi_{\mathrm p})  \right).   \ee
A comparison with~\eqref{Xvp2} leads to an exact temporal differential, which
when integrated yields a spatial function. Setting this function to zero
fixes the residual gauge freedom in the synchronous-comoving gauge and leads to a
one-to-one gauge invariant relationship with the total matter gauge
determined by
\be {}^{(1)}\!{\bf Z}_{{\mathrm s},{\mathrm v}}= -\sfrac23 m^{-2}x
{}^{(1)}\!\Psi_{\mathrm p}.  \label{Xvp3} \ee
Using~\eqref{Xvp1} and~\eqref{Xvp3} it follows from~\eqref{delta_transf} that
\be {}^{(1)}\!{\bdelta}_\mathrm{s} = {}^{(1)}\!{\bdelta}_\mathrm{v},\qquad
{}^{(2)}\!{\bdelta}_\mathrm{s} = {}^{(2)}\!{\bdelta}_\mathrm{v} - \sfrac43
xm^{-2}({\bf D}^i{\bdelta}_\mathrm{v}) ({\bf D}_i \Psi_{\mathrm p}).
\label{bdelta_2s} \ee

\end{appendix}


\begin{thebibliography}{99}

\bibitem{baretal04b} N.~Bartolo, E.~Komatsu, S.~Matarrese and A.~Riotto.
\newblock Non-Gaussianity from inflation: theory and observations,
\newblock  Physics\ Reports\ {\bf 402}, 103-266 (2004).

\bibitem{baretal10b} N.~Bartolo, S.~Matarrese and A.~Riotto.
\newblock Non-Gaussianity and the Cosmic Microwave Background Anisotropies.
\newblock  Advances in Astronomy, {\bf 2010}, 157079, (2010), arXiv:1001.3957 [astro-ph.CO].

\bibitem{pitetal10} C.~Pitrou, J-P.~Uzan,  and F.~Bernardeau.
\newblock The cosmic microwave background bispectrum from
the non-linear evolution of the cosmological perturbations.
\newblock JCAP\ {\bf1007}, 003 (2010).

\bibitem{matetal98} S.~Matarrese, S.~Mollerach and M.~Bruni.
\newblock Relativistic second-order perturbations of the Einstein-de Sitter
universe.
\newblock Phys.\ Rev.\ D {\bf 58}, 043504 (1998). 

\bibitem{baretal05} N.~Bartolo, S.~Matarrese and A.~Riotto.
\newblock Signatures of Primordial Non-Gaussianity in the Large-Scale Structure of
the Universe.
\newblock JCAP\ {\bf 0510}, 010 (2005).

\bibitem{tom05} K.~Tomita.
\newblock Relativistic second-order perturbations of
nonzero-$\Lambda$ flat cosmological models and CMB anisotropies.
\newblock Phys.\ Rev.\ D\ {\bf 71}, 083504 (2005).

\bibitem{baretal06}	N.~Bartolo, S.~Matarrese and A.~Riotto.
\newblock The full second-order radiation transfer function for large-scale CMB
anisotropies.
\newblock JCAP\ {\bf 0605}, 010 (2006).

\bibitem{baretal10} N.~Bartolo, S.~Matarrese, O.~Pantano and A.~Riotto.	
\newblock Second-order matter perturbations in a $\Lambda$CDM cosmology and
non-Gaussianity.
\newblock Class.\ Quantum\ Grav.\ {\bf 27}, 124009 (2010).

\bibitem{bruetal14} M.~Bruni, J.~C.~Hidalgo, N.~Meures and D.~Wands.
\newblock Non-Gaussian initial conditions in $\Lambda$CDM: Newtonian,
relativistic and primoridial contributions.
\newblock  Astrophys.\ J.\ {\bf 785}, 2 (2014) arXiv:1307.1478v1.

\bibitem{chalew11} A.~Challinor and A.~Lewis.
\newblock Linear power spectrum of observed source number counts.
\newblock Phys.\ Rev.\ {\bf D84} 043516 (2011).
\newblock DOI: http://dx.doi.org/10.1103/PhysRevD.84.043516

\bibitem{bondur11} C.~Bonvin and R.~Durrer.
\newblock What galaxy surveys really measure.
\newblock Phys.\ Rev.\ {\bf D84} 063505 (2011).
\newblock DOI: http://dx.doi.org/10.1103/PhysRevD.84.063505

\bibitem{jeoetal12} D.~Jeong, F.~Schmidt, and C.~M.~Hirata.
\newblock Large-scale clustering of galaxies in general relativity.
\newblock Phys.\ Rev.\ {\bf D85} 023504 (2012).
\newblock DOI: http://dx.doi.org/10.1103/PhysRevD.85.023504

\bibitem{bruetal12} M.~Bruni, R.~ Crittenden,  K.~ Koyama,  R.~ Maartens,  C.~ Pitrou and D.~ Wands.
\newblock Disentangling non-Gaussianity, bias and GR effects in the galaxy distribution.
\newblock Phys.\ Rev.\ {\bf D85} 041301 (2012).
\newblock  DOI: http://dx.doi.org/10.1103/PhysRevD.85.041301

\bibitem{yooetal12} J.~Yoo, N.~Hamaus, U.~Seljak and M.~Zaldarriaga.
\newblock Going beyond the Kaiser redshift-space distortion formula: A full general relativistic
account of the effects and their detectibility in galaxy clustering.
\newblock Phys.\ Rev.\ {\bf D86} 063514 (2012).
\newblock DOI: http://dx.doi.org/10.1103/PhysRevD.86.063514

\bibitem{beretal14} D.~Bertacca, R.~Maartens and C.~ Clarkson.
\newblock Observed galaxy number counts on the lightcone up to second order: I. Main result.
\newblock arXiv:1405.4403 (2014).

\bibitem{uggwai13} C.~Uggla and J.~Wainwright.
\newblock Asymptotic analysis of perturbed dust cosmologies to second order.
\newblock Gen.\ Rel.\ Grav.\ {\bf 45} 1467 (2013) doi:10.1007/s10714-013-1559-0.

\bibitem{hwanoh06} J-C.~Hwang and H.~Noh.
\newblock Second order perturbations of a zero-pressure cosmological medium: comoving vs. synchronous-comoving gauge.
\newblock Phys.\ Rev.\ {\bf D73} 044021 (2006).

\bibitem{hwaetal12} J-C.~Hwang, H.~Noh, and  J-O.~Gong.
\newblock Second order solutions of cosmological perturbation in the matter dominated era.
\newblock Astrophys.\ J.\ {\bf 752}, 50 (2012) doi:10.1088/0004-637X/752/1/50.

\bibitem{uggwai12} C.~Uggla and J.~Wainwright.
\newblock Scalar Cosmological Perturbations.
\newblock Class.\ Quantum\ Grav.\ {\bf 29} 105002 (2012) doi:10.1088/0264-9381/29/10/105002.

\bibitem{malwan09} K.~A.~Malik and D.~Wands.
\newblock Cosmological perturbations.
\newblock Physics\ Reports\ {\bf 475}, 1-51(2009).

\bibitem{hwanoh99} J-C.~Hwang and H.~Noh.
\newblock Relativistic hydrodynamic cosmological perturbations.
\newblock  Gen.\ Rel.\ Grav.\ {\bf 31} 1131 (1999).

\bibitem{wandsetal00} D.~Wands, K.~A.~Malik, D.~H.~Lyth, and A.~R.~Liddle.
\newblock A new approach to the evolution of cosmological perturbations on large
scales.
\newblock Phys.\ Rev.\ D\ {\bf 62}, 043527 (2000).

\bibitem{malwan04} K.~A.~Malik and D.~Wands.
\newblock Evolution of second order cosmological perturbations.
\newblock Class.\ Quantum\ Grav.\ {\bf 21}, L65 (2004)

\bibitem{lanver05}  D.~Langlois and F.~Vernizzi.
\newblock Conserved nonlinear quantities in cosmology.
\newblock  Phys.\ Rev.\ D {\bf 72}, 103501 (2005).

\bibitem{baretal04a} N.~Bartolo, S.~Matarrese and A.~Riotto.
\newblock Gauge-Invariant Temperature Anisotropies and Primordial Non-Gaussianity,
\newblock  Phys.\ Rev.\ Lett.\ {\bf 93}, 231301 (2004).

\bibitem{uggwai13c} C.~Uggla and J.~Wainwright.
\newblock Simple expressions for second order density perturbations in standard cosmology.
\newblock  Class.\ Quantum\ Grav.\ {\bf 31} 105008 (2014), arXiv:1312.1929.

\bibitem{uggwai13b} C.~Uggla and J.~Wainwright.
\newblock A simplified structure for the second order cosmological perturbation equations.
\newblock Gen.\ Rel.\ Grav.\ {\bf 45}, 643 (2013) doi:10.1007/s10714-012-1492-7.

\bibitem{goowai85} S.~W.~Goode and J.~Wainwright.
\newblock  Isotropic singularities in cosmological models.
\newblock Class.\ Quantum\ Grav.\ {\bf 2},99 (1985).

\bibitem{bruetal97} M.~Bruni, S.~Matarrese, S.~Mollerach, S.~Sonego.
\newblock Perturbations of spacetime: gauge transformations and gauge-invariance at second order and
beyond.
\newblock Class.\ Quantum\ Grav.\ {\bf 14}, 2585 (1997).

\bibitem{chretal11} A.~J.~Christopherson, K.~A.~Malik, D.~R.~Matravers and
    K.~Nakamura.
\newblock  Comparing different formulations of nonlinear perturbation theory.
\newblock Class.\ Quantum\ Grav.\ {\bf 28}, 225024 (2011).

\bibitem{hidetal13} J.~C.~Hidalgo,  A.~J.~Christopherson and K.~A.~Malik.
\newblock  The Poisson equation at second order in relativistic cosmology.
\newblock arXiv:1303.3074v2.

\bibitem{uggwai11} C.~Uggla and J.~Wainwright.
\newblock  Cosmological perturbation theory revisited.
\newblock Class.\ Quantum\ Grav.\ {\bf 28} 175017 (2011) doi:10.1088/0264-9381/28/17/175017.

\bibitem{nak06} K.~Nakamura.
\newblock Gauge-invariant Formulation  of the Second-order Cosmological  Perturbations.
\newblock Phys.\ Rev.\ D\ {\bf 74}, 101301 (2006).

\bibitem{nak07} K.~Nakamura.
\newblock Second Order Gauge Invariant Cosmological Perturbation Theory.
\newblock Prog.\ Theor.\ Phys.\ {\bf 117}, 17 (2007).


\end{thebibliography}
\end{document}